\documentclass[11pt]{article}
\pdfoutput=1
\usepackage{authblk}
\usepackage{palatino}
\usepackage{amsmath}
\usepackage{amsthm}
\usepackage{amssymb}
\usepackage{boxedminipage}
\usepackage[mathcal]{euscript}
\usepackage{graphics}
\usepackage{graphicx}
\usepackage{caption}
\usepackage{multirow}
\usepackage{url}

\newtheorem{theorem}{\bf Theorem}[section]

\newtheorem{lemma}[theorem]{\bf Lemma}

\newcommand{\tenT} {{\mathcal T}}

\newcommand{\tenZ} {{\mathcal Z}}
\newcommand{\calG} {{\mathcal G}}
\newcommand{\calE} {{\mathcal E}}

\newcommand{\Rnp} {{\tt R}_n^p}

\newcommand{\kron} {{\sf Kron}}
\newcommand{\Real} {\mathbb{R}}
\newcommand{\khat} {\widehat{K}}
\newcommand{\Lhat} {\widehat{L}}
\newcommand{\xin} {\overrightarrow{x}_{\rm in}}
\newcommand{\xout} {\overrightarrow{x}_{\rm out}}
\newcommand{\yin} {\overrightarrow{y}_{\rm in}}
\newcommand{\yout} {\overrightarrow{y}_{\rm out}}

\newcommand{\gcore} {{\mathcal G}}

\newcommand{\slc} {{\sf Slice}}
\newcommand{\matF} {\mathbf{F}}
\newcommand{\tmatF} {\mathbf{F}^T}
\newcommand{\matA} {\mathbf{A}}

\newcommand{\newF} {\widetilde{\matF}}
\newcommand{\tnewF} {\widetilde{\matF}^T}
\newcommand{\newg} {\tilde{\gcore}}

\newcommand{\ucar} {U{\c{c}}ar}
\newcommand{\mnuf}[1] {\mathbf{#1}_{(n)}}

\newcommand{\val} {{\sf val}}
\newcommand{\eat}[1] {}

\newcommand{\Rnsum} {{\tt R}_n^{\tt sum}}
\newcommand{\Rnmax} {{\tt R}_n^{\tt max}}
\newcommand{\Enmax} {{\tt E}_n^{\tt max}}
\newcommand{\Emax} {{\tt E}^{\tt max}}

\newcommand{\share} {{\tt num}}

\newcommand{\avgE} {\left \lceil |\calE|/P\right\rceil}
\newcommand{\avgL} {\left\lceil L_n/P\right\rceil}
\newcommand{\lmt} {{\tt limit}}
\newcommand{\that} {\widehat{t}}
\newcommand{\hhat} {\widehat{h}}
\newcommand{\phat} {\widehat{p}}
\newcommand{\ghat} {\widehat{g}}
\newcommand{\calS} {{\mathcal S}}
\newcommand{\contr} {{\sf contr}}
\newcommand{\hyperg} {{\sf HyperG}}
\newcommand{\Lite} {{\sf Lite}}
\newcommand{\mediumg} {{\sf MediumG}}
\newcommand{\coarseg} {{\sf CoarseG}}

\begin{document}
\title{On Optimizing Distributed Tucker Decomposition for Sparse Tensors}
\author[1]{Venkatesan T. Chakaravarthy} 
\author[1]{Jee W. Choi} 
\author[1]{Douglas J. Joseph} 
\author[2]{Prakash Murali\footnote{Research conducted while the author was at IBM Research.}} 
\author[1]{Shivmaran S. Pandian} 
\author[1]{Yogish Sabharwal} 
\author[1]{Dheeraj Sreedhar}

\affil[1]{
	IBM Research
	\authorcr
	\it{\{vechakra,prakmura,ysabharwal,dhsreedh\}@in.ibm.com}\\
	\authorcr
	\it{\{jwchoi,djoseph,xliu\}@us.ibm.com}
}

\affil[2]{
	Princeton University
	\authorcr
	{\it pmurali@cs.princeton.edu}
}
\date{~}

\maketitle     
\begin{abstract}
The Tucker decomposition generalizes the notion of Singular Value Decomposition (SVD) to tensors, 
the higher dimensional analogues of matrices.
We study the problem of constructing the Tucker decomposition of sparse tensors on distributed memory systems
via the HOOI procedure, a popular iterative method.
The scheme used for distributing the input tensor among the processors (MPI ranks)
critically influences the HOOI execution time.
Prior work has proposed different distribution schemes: 
an offline scheme based on sophisticated hypergraph partitioning
method and simple, lightweight alternatives that can be used real-time.
While the hypergraph based scheme typically results in faster HOOI execution time,
being complex, the time taken for determining the distribution is an
order of magnitude higher than the execution time of a single HOOI iteration.  
Our main contribution is a lightweight distribution scheme, which achieves the best of both worlds.
We show that the scheme is near-optimal on certain fundamental metrics
associated with the HOOI procedure and as a result, near-optimal on the computational load (FLOPs).
Though the scheme may incur higher communication volume, 
the computation time is the dominant factor and as the result, the scheme
achieves better performance on the overall HOOI execution time.
Our experimental evaluation on large real-life tensors (having up to 4 billion elements)
shows that the scheme outperforms the prior schemes on the HOOI
execution time by a factor of up to 3x.
On the other hand, its distribution time
is comparable to the prior lightweight schemes and is typically lesser than the
execution time of a single HOOI iteration.
\end{abstract}

\newpage
\section{Introduction}
\begin{figure}[t]
	\centering
	\includegraphics[width=3.0in]{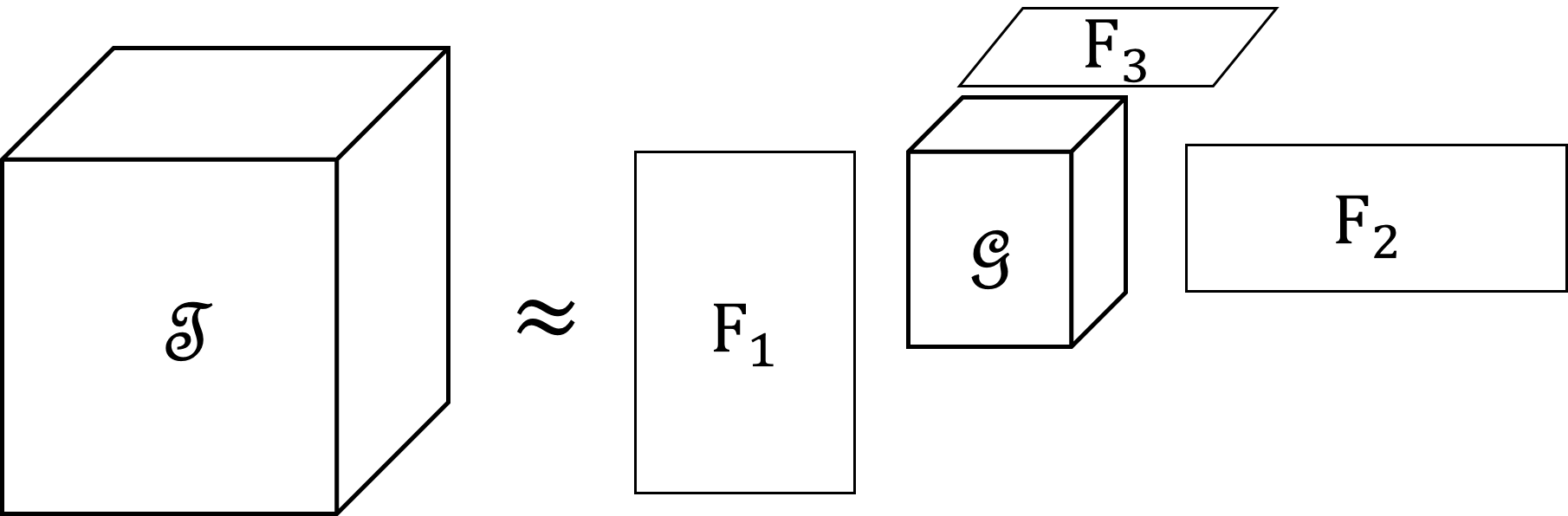}
	\caption{
		Illustration for the Tucker decomposition in $3$-D. $\calG$ is the core tensor, and $F_1$, $F_2$ and $F_3$ are the factor matrices.
	}
	\label{fig:tucker}
\end{figure}

Tensors are the higher dimensional analogues of matrices that are useful in representing data in three or higher dimensions.
Different tensor decompositions have been proposed, among which the two most prominent are 
the Tucker decomposition \cite{tucker} and CP (canonical polyadic) decomposition \cite{cp-original1,cp-original2}.
The Tucker decomposition represents high-rank data in the form of a low-rank structure:
given an $N$-dimensional input tensor $\tenT$, the decomposition (approximately) expresses $\tenT$
as the product of a small $N$-dimensional core tensor, and a set of $N$ factor matrices. 
It can be viewed as a generalization of the SVD (Singular Value Decomposition) to higher dimensions.
While the factor matrices represent the most significant information along the different dimensions,
the core captures the interaction among them. Figure \ref{fig:tucker} provides a pictorial depiction in $3$-D.
The CP decomposition generalizes rank factorization and can be viewed as a constrained form of Tucker decomposition, wherein the core is a diagonal tensor 
with uniform length across all the dimensions.

The Tucker decomposition has been used in performing tasks such as data compression and principal component analysis.
It finds application in diverse domains from signal processing \cite{signal-app} to text analytics \cite{text-app}. 
We refer to the survey by Kolda and Bader \cite{survey} for a detailed discussion on Tucker decomposition and its applications.
The decomposition has been well-studied in sequential, shared memory, map-reduce and distributed settings, 
for both dense and sparse tensors \cite{kolda-ipdps,matlab, baskaran,our-dense,haten2,ucar-icpp,kolda-icdm,dms-ipdps,smith-ipdps17,europar}.

\paragraph{HOOI Procedure.}
The HOOI (Higher Order Orthogonal Iterator) procedure \cite{hooi} provides a popular iterative method for constructing the Tucker decomposition of a given tensor.
The procedure transforms any given decomposition to a more refined decomposition
and it is usually invoked multiple times until a suitable convergence criterion is attained.
The procedure is bootstrapped with an initial decomposition obtained via
methods such as the HOSVD (Higher Order SVD) algorithm \cite{hosvd}; alternatively, a random set of factor matrices can also be used.
Given a decomposition, the HOOI procedure (a single invocation) 
constructs the new decomposition in $N$ iterations, 
wherein each iteration involves a TTM-Chain (Tensor-times-matrix chain) operation, followed by an SVD step. 
Finally, the newly constructed factor matrix rows are communicated among the processors
to be used in the subsequent HOOI invocation.

Our goal is to develop an efficient implementation of the Tucker decomposition for sparse tensors on distributed memory systems.
We build on a prior framework of Kaya and {\ucar} \cite{ucar-icpp},
which provides mechanisms for implementing the TTM and SVD operations in a distributed manner.
The execution time of the HOOI procedure critically depends on the scheme used for distributing the input tensor among the processors (MPI ranks).

\paragraph{Prior Schemes.}
We consider distribution schemes proposed for Tucker decomposition, as well as the related CP decomposition.
The schemes can be classified into three types:
(i) coarse-grained schemes \cite{defacto, coarse2, ucar-icpp} that partition the tensor into large chunks (sub-tensors)
and assign the chunks to the processors;
(ii) fine-grained schemes that assign individual tensor elements \cite{ucar-icpp};
(iii) a medium grained scheme \cite{dms-ipdps} that strikes a balance between the two.
Among the above methods, hypergraph partitioning (a fine grained schme) \cite{ucar-icpp}
typically offers the best HOOI execution time.
However, hypergraph partitioning is expensive and the time taken for distributing the tensor is significantly higher than
the execution time of a single HOOI invocation.
On the other hand, the other schemes are real-time, lightweight procedures
with much faster distribution time (comparable to HOOI execution).

\paragraph{Our Contributions.}
Our main goal is to demonstrate that 
high performance on the HOOI execution time can be achieved via lightweight schemes:
\begin{itemize}
\item
We present a lightweight distribution scheme called {\Lite} that is easy to implement and parallelize.
\item
We define certain fundamental metrics (implicit in the prior work) associated with the HOOI procedure 
and prove that {\Lite} is near-optimal on all these metrics.
As a result, the scheme is near-optimal on computational load, load balance
and communication volume associated with the TTM and the SVD components.
\item
{\Lite} outperforms the prior schemes on real-life tensors in terms of the HOOI execution time.
{\Lite} may incur higher overall communication volume, because of higher factor matrix data transfer. 
Nevertheless, in contrast to the CP decomposition, the computation time is the dominant factor in the Tucker decomposition
and as a result, {\Lite} achieves better HOOI execution time.
\end{itemize}
We present a detailed experimental study evaluating the different schemes
over a benchmark of large real-life tensors having up to $4$ billion elements. 
The results show that the new scheme achieves the best of both worlds:
\begin{itemize}
\item
On HOOI execution time, {\Lite} outperforms the hypergraph based scheme by a factor of up to $4$x.
Taking the best of prior schemes in each test case, the gain is upto a factor of $3$x.
HOOI scales well under the scheme: on MPI ranks from $32$ to $512$,
the speedup is in the range of $8.7$x to $15.5$x.
\item 
	{\Lite} distributes tensors with billions of elements in real-time,
with its distribution time comparable to the prior lightweight schemes and a single HOOI invocation. 
\end{itemize}
%We also consider the memory requirements of the HOOI procedure under the different schemes.
%While the hypergraph based scheme stores only a single copy of the tensor (distributed among the processors),
%the new scheme stores $N$ copies, each distributed in a manner suitable along each dimension (mode).
%However, due to improvements in the fundamental metrics, the HOOI procedure under the new scheme 
%incurs lesser memory usage in other components. 
%Consequently, the overall memory footprint under {\Lite} is lesser than that of the hypergraph based scheme.
%Finally, being lightweight, {\Lite} is able to handle large tensors with billions of elements,
%whereas we could not obtain the corresponding hypergraph partitioning.

\paragraph{Related Work.}
Tucker decomposition has been studied under various settings.
In the case of sparse tensors, the direct evaluation of TTM-Chain via computing intermediate tensors
leads to memory blowup. To address the issue, Kolda and Sun \cite{kolda-icdm} proposed a memory efficient approach (MET)
and Baskaran \cite{baskaran} developed semi-dense structures.
Recently, Smith and Karypis \cite{europar} used the compressed sparse fiber representation (CSF)
to reduce the computational load, while limiting the memory blowup.
The above implementations target sequential and shared-memory settings.
Kaya and {\ucar} \cite{ucar-icpp} presented the first distributed memory implementation;
we build on their framework.
The TTM component of the framework is a special case of the MET approach, wherein no intermediate tensors are computed. 

For the Tucker decomposition of dense tensors, MATLAB \cite{matlab}, 
single-machine \cite{zhou} and distributed \cite{kolda-ipdps,our-dense} implementations have been proposed.
Prior work has also studied the Tucker decomposition on the MapReduce platform \cite{haten2}.
Other tensor decompositions such as CP factorization have been explored as well (e.g.,\cite{cp1,cp2,cp3,dms-ipdps,hyper1}).

\begin{figure}[t]
\begin{center}
\begin{boxedminipage}{\hsize}
\begin{tabbing}
xx\=xx\=xx\=xx\=xx\=xx\=xx\=\kill
\textbf{Input:} A tensor $\tenT$ and a decomposition $\{\gcore, \matF_1, \matF_2, \ldots, \matF_N\}$ \\
\> Size of tensor $\tenT$: $L_1\times L_2\times \cdots\times L_N$\\
\> Size of core $\gcore$: $K_1\times K_2\times \cdots K_N$,\\
\> Size of factor matrix $\matF_n$: $L_n\times K_n$\\
\textbf{Output:} A new refined decomposition $\{\newg, \newF_1, \newF_2, \ldots, \newF_N\}$ \\
\> Size of core $\newg$ and factor matrices $\newF_n$: same as input.\\
\textbf{Procedure:}\\
For each mode $n$ from $1$ to $N$\\
\> {\sf TTM-Chain:} Perform TTM along all the modes, except $n$.\\
\> ~~$\tenZ \leftarrow \tenT \times_1 \tmatF_1\times \cdots \times_{n-1} \tmatF_{n-1}\times_{n+1} \tmatF_{n+1}\times \cdots \times_N \tmatF_N$.\\
\> {\sf SVD:} $\newF_n \leftarrow$ leading $K_n$ left singular vectors of $\mnuf{Z}$\\
{\sf New core:} $\newg \leftarrow \tenT\times_1 \tnewF_1\times \cdots \times_N \tnewF_N$\\
\textbf{return} $\{\newg, \newF_1, \newF_2, \ldots, \newF_N\}$.
\end{tabbing}
\end{boxedminipage}
\end{center}
\caption{HOOI Procedure}
\label{fig:hooi}
\end{figure}

\section{Tucker Decomposition}
In this section, we briefly describe tensor concepts pertinent to our discussion,
and present the HOOI procedure. 

\subsection{Tensors}
\paragraph{Fibers.}
Tensors are multi-dimensional arrays.
Consider an $N$-dimensional tensor $\tenT$ of size $L_1\times L_2\times \cdots \times L_N$.
The elements of $\tenT$ can be canonically indexed by a coordinate vector $\langle l_1, l_2, \ldots, l_N\rangle$,
where each $l_j$ belongs to $[1, L_j]$.
For $1\leq n\leq N$, a {\em mode-$n$ fiber} refers to the vector obtained by fixing all the coordinates except $n$,
i.e., $\langle l_1, \ldots, l_{n-1}, *, l_{n+1}, \ldots l_N\rangle$.
These fibers have length $L_n$ and the number of fibers is $\Lhat_n = \Pi_{j\neq n} L_j$.
Each fiber can be identified by an index $\langle l_1, l_2, \ldots, l_{n-1}, l_{n+1},\ldots, l_N\rangle$.
In the analogous case of matrices, two types of fibers can be found: row vectors and column vectors.

\paragraph{Tensor Unfolding.}
A {\em mode-$n$ unfolding} refers to the matrix of size $L_n \times \Lhat_n$ obtained by arranging the mode-$n$ fibers as the columns.
We adopt a standard lexicographic ordering of the fibers, wherein
the fiber indexed $\langle l_1, l_2, \ldots, l_{n-1}, l_{n+1},\ldots, l_N\rangle$
is placed in the column numbered $\Sigma_{j \neq n} l_j \large(\Pi_{i \neq n, i < j} L_i\large)$.
The matrix is denoted as $\mnuf{T}$.

\paragraph{Tensor-Times-Matrix Multiplication (TTM).}
The tensor $\tenT$ can be multiplied along mode-$n$ by any matrix $\matA$ provided $\matA$ has size $K\times L_n$ (for some $K$)
and the operation is denoted $\tenZ = \tenT \times_n \matA$.
Conceptually, the operation applies the linear transformation $\matA$ to all the mode-$n$ fibers.
It is realized via the matrix-matrix multiplication $\matA \times \mnuf{T}$, 
and taking the output matrix to be the mode-$n$ unfolding of $\tenZ$.
The output tensor retains the same length along all modes, except mode $n$, where its length becomes $K$,
(i.e., $\tenZ$ has size $L_1\times \cdots \times L_{n-1}\times K\times L_{n+1}\times \cdots \times L_N$).

\paragraph{TTM-Chain.}
The {\em TTM-chain} operation refers to multiplying $\tenT$ along multiple distinct modes $S=\{n_1, n_2, \ldots, n_r\}$
by matrices $\matA_1, \matA_2, \ldots, \matA_r$, where $\matA_j$ has size $K_j \times L_{n_j}$.
The output is a tensor $\tenZ$ whose length 
remains $L_j$, for all modes $j\not\in S$ and changes to $K_j$, for all modes $j\in S$.
We denote the operation as $\tenZ = \tenT \times_{n_1}\matA_1 \times \cdots\times_{n_r} \matA_{n_r}$.
The operation is commutative, namely the $r$ TTMs can be performed in any order \cite{hooi}..

\subsection{HOOI Procedure}
The Tucker decomposition of $\tenT$ approximately represents the tensor
as the product of a small {\em core tensor} $\gcore$ and a set of {\em factor matrices} $\matF_1, \matF_2, \ldots, \matF_N$:
\[
\tenT \approx \tenZ = \gcore\times_1 \matF_1\times_2 \matF_2 \times \cdots \times_N \matF_N.
\]
The core is of size $K_1\times K_2\times \cdots \times K_N$, with each $K_n \leq L_n$, which are user-specified.
Each factor matrix $\matF_n$ has size $L_n \times K_n$,
We write the decomposition as $\{\gcore;\matF_1,\matF_2,\ldots,\matF_N\}$.

The HOOI procedure \cite{hooi} is a popular method that generalizes SVD to higher order tensors
and produces a Tucker decomposition with the additional property that the factor matrices are orthonormal.
The procedure takes as input any decomposition $\{\gcore, \matF_1, \matF_2, \ldots, \matF_N\}$ 
and produces a new, more refined decomposition $\{\newg, \newF_1, \newF_2, \ldots, \newF_N\}$.
It can be invoked multiple times to obtain better refinements
until a suitable convergence criterion is reached (such number of invocations fixed a priori).
The process must be bootstrapped by providing an initial decomposition,
which we can find using methods such as HOSVD \cite{hosvd};
alternatively, random factor matrices can also be used.

The HOOI procedure, a single invocation, is shown in Figure \ref{fig:hooi}.
For computing each new factor matrix $\newF_n$, the procedure utilizes the alternating least squares paradigm and works in two steps.
First, it performs a TTM-chain operation by skipping mode $n$ and multiplying $\tenT$ by the transposes of all the other factor matrices $\matF_j$ (with $j\neq n$)
and obtains a tensor $\tenZ$. 
The tensor $\tenZ$ has length compressed from $L_j$ to $K_j$ along all modes $j\neq n$.
The mode-$n$ unfolding of $\tenZ$, denoted $\mnuf{Z}$, is a matrix of size $L_n\times \khat_n$,
where $\khat_n = \Pi_{j\neq n} K_j$.
In the next step, the procedure performs an SVD on $\mnuf{Z}$ and obtains the leading $K_n$ singular vectors of $\mnuf{Z}$.
These singular vectors are arranged in the form of columns to obtain the new factor matrix $\newF_n$.
We refer to the matrix $\mnuf{Z}$ as the {\em penultimate matrix}, since it is one TTM short of a full TTM chain.

The new core tensor is computed in the last step.
However, we can see that the refinement procedure only involes the factor matrices, 
and so it is not necessary to compute the core in each invocation. 
Instead, it suffices to compute the core only once after all the invocations are completed.
Consequently, we focus on optimizing the computation of the TTM-chain and the SVD components.

\section{Distributed Framework}
\label{sec:outline}
As mentioned in the introduction, we build on the distributed framework of Kaya and {\ucar} \cite{ucar-icpp}.
Here, we present an outline of the framework focusing on the aspects critical for our discussion. 

The input sparse tensor $\tenT$ is represented in the coordinate format.
Let $\calE$ denote the set of all non-zero elements. 
Each element $e\in \calE$ is represented by a coordinate vector $(l_1, l_2, \ldots, l_N)$ (where each $l_n\in [1,L_n]$) and a value $\val(e) \in \Real$.
Consider a distributed setting consisting of $P$ processors (MPI ranks), numbered $0,1, \ldots, P-1$.

The HOOI procedure involves of $N$ iterations.  Consider the computation along any mode $n\in [1,N]$,
consisting of a TTM-Chain operation that generates the penultimate matrix $\mnuf{Z}$ of size $L_n\times \khat_n$,
followed by an SVD operation on the matrix.
In order to evaluate the TTM-Chain in a distributed manner, 
the framework uses a reformulation via the Kronecker product. 

\begin{figure}[t]
	\center
	\includegraphics[width=4in]{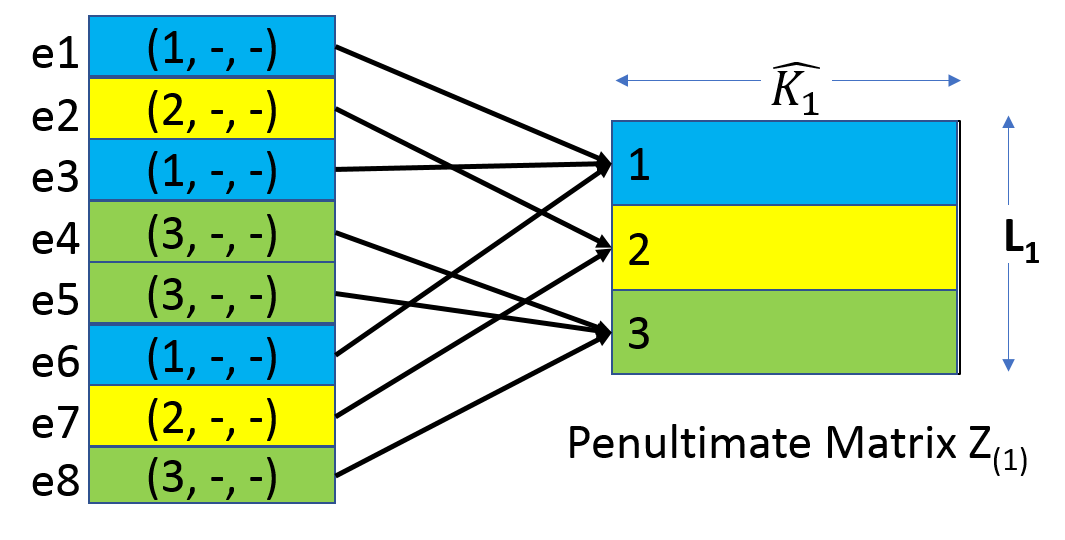}
	\caption{TTM-Chain reformulation}
	\label{fig:seq-kron}
\end{figure}

\paragraph{Reformulation.}
We partition the elements into groups based on the $n^{th}$ coordinate, called {\em slices}:
for each $l\in [1, L_n]$, define $\slc_n^l$ as the set of elements having the $n^{th}$ coordinate as $l$.
The reformulation is based on the observation that any row $\mnuf{Z}[l,:]$ is determined only by the contributions from the elements in $\slc_n^l$.
Figure \ref{fig:seq-kron} provides an illustration using a $3$-D tensor with eight elements,
by considering the TTM-Chain operation along the first mode $(n=1)$. 
In this example, $L_1 = 3$ and so, there are three slices: $\slc_1^1 = \{e_1, e_3, e_6\}$, $\slc_1^2 = \{e_2, e_7\}$ and $\slc_1^3 = \{e_4, e_5, e_8\}$.
The rows to which the elements contribute are shown by arrows.
Each slice and the corresponding row are assigned the same color.

While the row to which $e$ contributes is determined by its $n^{th}$ coordinate,
the contribution is determined by the other $(N-1)$ coordinates of $e$.
The contribution, denoted $\contr_n(e)$, is a vector of length $\khat_n$, the same as the length of rows of $\mnuf{Z}$.
It is computed via the Kronecker (or outer) product 
of the rows indexed by the above $(N-1)$ coordinates in the corresponding factor matrices.
We vectorize the resultant $(N-1)$-dimensional tensor and scale by $\val(e)$ to get $\contr_n(e)$;
the details are provided in the Appendix.
The reformulation states that for any row-index $l\in [1,L_n]$, 
the row $l^{th}$is given by: 
\begin{eqnarray}
	 \mnuf{Z}[l,:] = \sum_{e\in \slc_n^l} \contr_n(e).
	 \label{eqn:reform}
\end{eqnarray}

\begin{figure}[t]
	\center
	\includegraphics[width=6in]{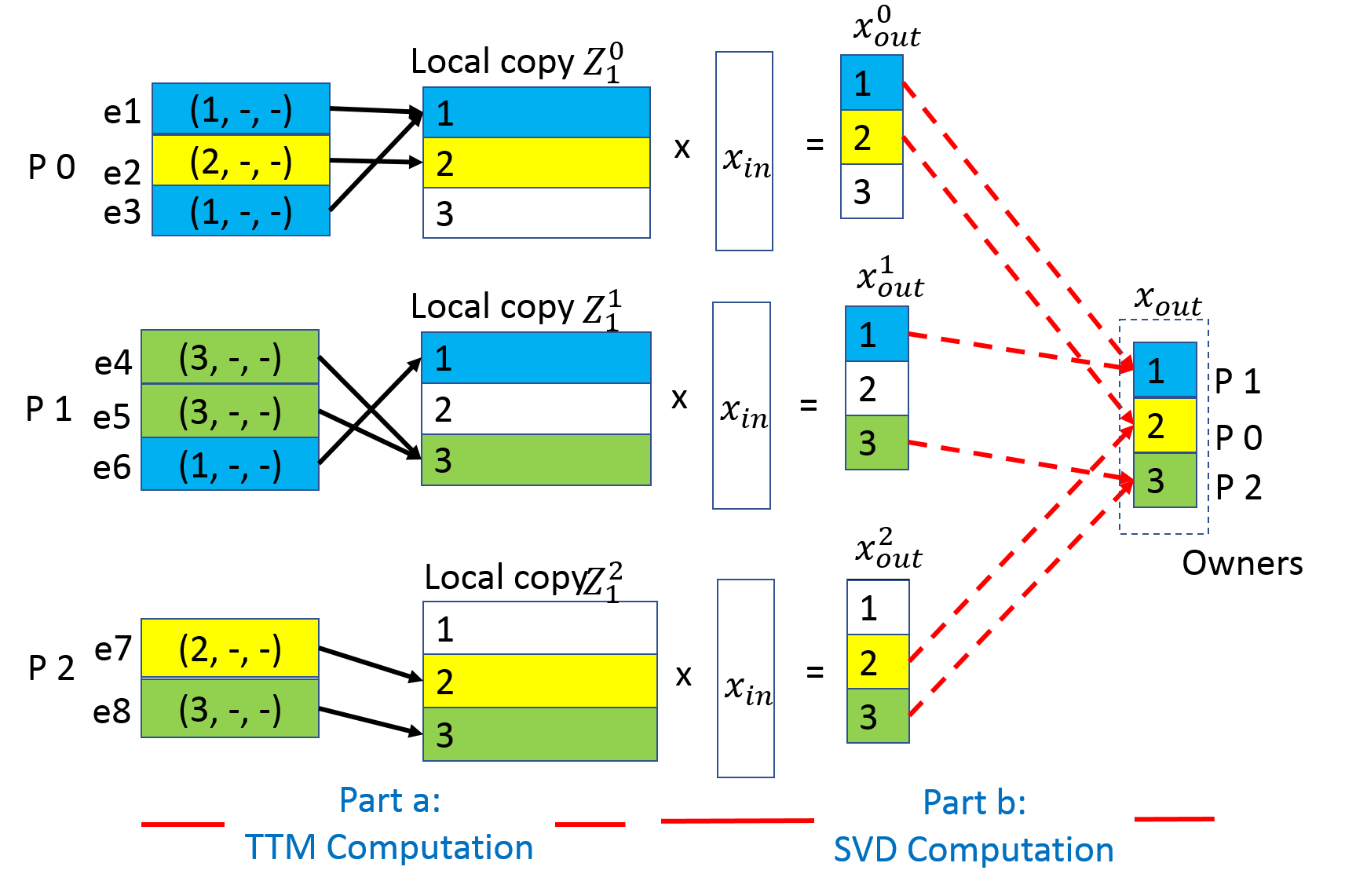}
	\caption{Framework illustration}
	\label{fig:framework}
\end{figure}

\paragraph{TTM Component.}
We distribute the input tensor using a {\em distribution policy} (a mapping) $\pi:\calE \rightarrow [0, P-1]$ that 
assigns each element $e$ to a processor $p=\pi(e)$, called the {\em owner} of $e$.
Equivalently, the policy partitions the set of elements $\calE$ into $P$ parts $\calE^0, \calE^1, \ldots, \calE^P$,
where $\calE^p$ denotes the set of elements assigned to the processor $p$.
Given a policy $\pi$, each processor $p$ computes a local copy of the penultimate matrix $\mnuf{Z}^p$ by considering 
only the contributions made by the elements owned by it. 
Namely, we initialize $\mnuf{Z}^p$ to all $0$ and for each element $e\in \calE^p$,
we add the vector $\contr_n(e)$ to the row $\mnuf{Z}^p[l, :]$, where $l$ is the $n^{th}$ coordinate of $e$.
The global penultimate matrix $\mnuf{Z}$ is simply the sum of the local copies, i.e.,
$\mnuf{Z}$ is sum-distributed.

Part (a) of Figure \ref{fig:framework} provides an illustration over three processors using a simple policy 
$\pi$ that partitions the elements in a lexicographic manner:
$\calE^0 = \{e_1, e_2, e_3\}$, $\calE^1 = \{e_4, e_5, e_6\}$ and $\calE^2 = \{e_7, e_8\}$. 
The local copies are also shown.

In the above procedure, a processor may not contribute to all the rows of $\mnuf{Z}$.
For a row-index $l\in [1, L_n]$, we say that a processor $p$ {\em shares} $\slc_n^l$, if it owns at least one element from the slice,
i.e., $\calE^p\cap \slc_n^l\neq \emptyset$. The processor contributes only to the rows corresponding to the slices shared by it; the other rows are said to be empty. 
Let $\Rnp$ denote the number of slices shared by $p$, or equivalently, the number of rows to which $p$ contributes.
By omitting the empty rows, we can represent the local copy succinctly as a matrix of size $\Rnp \times \khat_n$.
Apart from reducing the memory footprint, the above truncation provides significant advantages in optimizing the subsequent SVD operation.
In Figure \ref{fig:framework}, the empty rows are colored white; here, $L_1=3$ and $\Rnp=2$ for all the processors.

\paragraph{SVD Component.}
While the matrix $\mnuf{Z}$ can be constructed explicitly by aggregating the local copies,
the approach may lead to high volume of communication. 
Instead, the framework performs the SVD operation directly over the local copies by employing the Lanczos bidiagonalization method \cite{slepc},
an iterative matrix-free procedure.  The Lanczos method can be explained using an {\em oracle} (i.e., query-answering) model.
The method works iteratively, wherein each iteration generates two query vectors,
a column vector $\xin$ and a row vector $\yin$, and our task is to evaluate the matrix-vector products 
$\xout = \mnuf{Z}\cdot \xin$ and $\yout = \yin \cdot \mnuf{Z}$
and return the answers $\xout$ and $\yout$ to the method.  

Regarding the first product, each processor computes the local answer $\xout^p = \mnuf{Z}^p\cdot \xin$.
These get aggregated by a global point to point reduction operation, as follows.
The framework uses a suitable {\em row-index mapping} $\sigma_n:[1,L_n] \rightarrow [0,P-1]$
that assigns each row-index $l$ to a processor $\sigma_n(l)$ called the owner of $l$.
The owner is chosen to be one among the processors sharing $\slc_n^l$.
The owners accumulate partial contributions received from the other processors. 
Thus, the global answer $\xout$ is output in a distributed manner according to $\sigma_n$.
In Figure \ref{fig:framework}, processors $1$, $0$ and $2$ are the owners and communication is shown by dashed arrows.

\begin{figure*}
	\centering
	\includegraphics[width=6in]{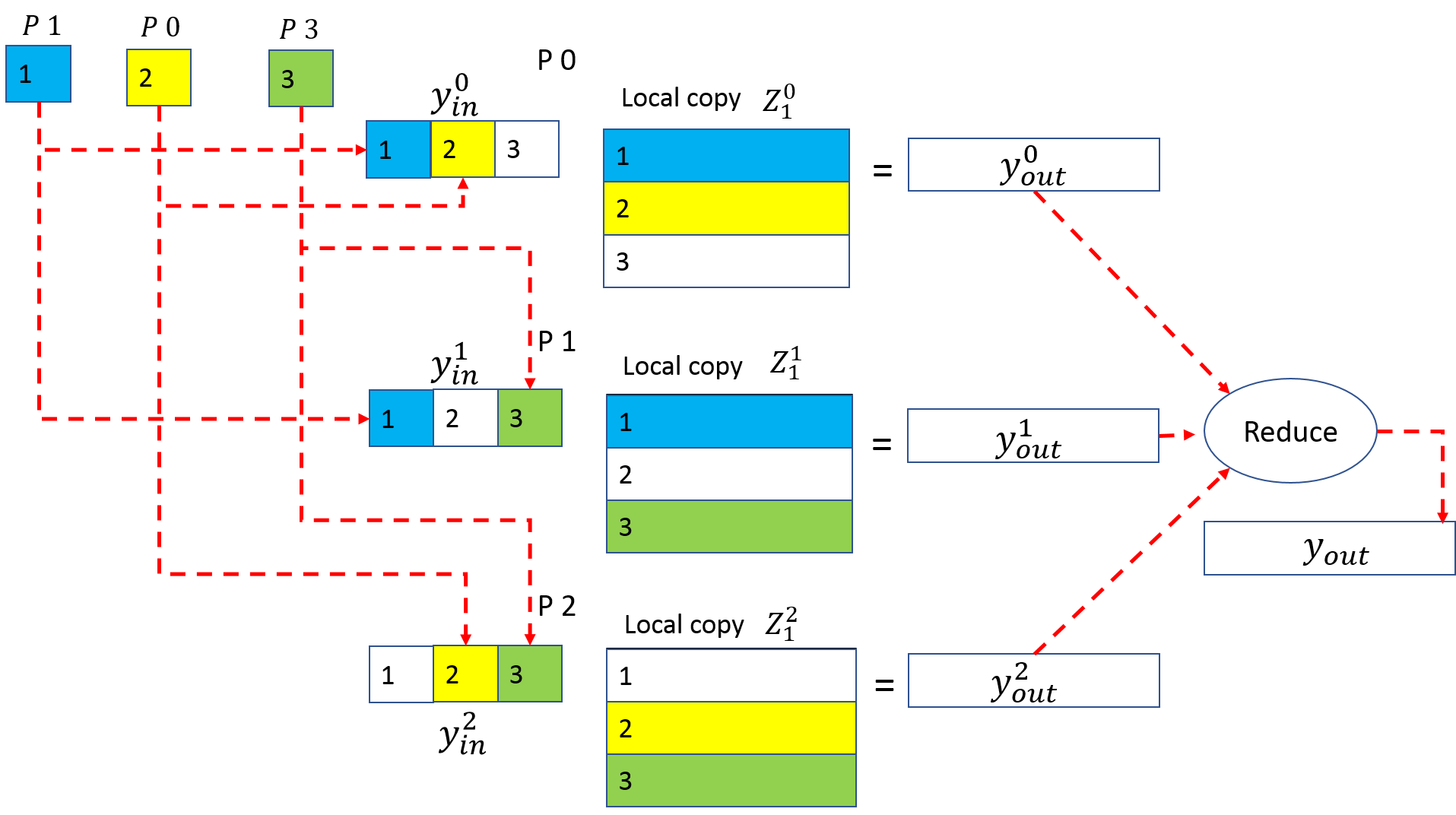}
	\caption{Vector-Matrix Product}
	\label{fig:mtv}
\end{figure*}

The second product $\yout = \yin \cdot \mnuf{Z}$ is executed in an analogous manner.
For each coordinate $l\in [1,L_n]$,
the owner $\sigma_n(l)$ sends the value $\yin(l)$ to all the processors sharing $\slc_n^l$.
Upon receiving the above values, each processor $p$ assembles a partial vector $\yin^p$ of length $\Rnp$
and performs the product $\yout^p = \yin^p \mnuf{Z}^p$. 
We obtain the answer vecotor $\yout$ by doing the summation $\yout = \sum_p \yout^p$ (via {\tt MPI\_Allreduce}).
Figure \ref{fig:mtv} illustrates the process. 

\paragraph{Factor Matrix Transfer.}
The Lanczos algorithm produces the factor matrix $\newF_n$ in a distributed manner, wherein each row $\newF_n[l,:]$
gets generated at the owner $\sigma_n(l)$. 
These rows are needed for the TTM computation of the next HOOI invocation. 
Towards that goal, the owner $\sigma_n(l)$ sends the row
$\newF_n[l,:]$ to all the processors that would require the row for the subsequent TTM computation.

\paragraph{Distribution Schemes.}
The execution time of the HOOI procedure critically depends on the choice of the distribution policy $\pi$,
since the policy determines the parameters of computational load (FLOPs), load balance and communication volume.
An efficient policy must optimize the above parameters with respect to the HOOI computation along all the $N$ modes. 
The task becomes easier, if we use $N$ distribution policies,
each customized with respect to the computation along a single mode.
We call such a sequence of $N$ policies $(\pi_1, \pi_2, \ldots, \pi_N)$ as a {\em distribution scheme}.
If a single policy $\pi$ is used across all modes, we refer to the scheme as {\em uni-policy scheme},
and the general case as {\em mulit-policy scheme}.
Uni-policy schemes need to store only a single copy of the input tensor (distributed among the processors),
whereas multi-policy schemes must store $N$ copies, one along each mode.
However, multi-policy schemes offer more flexibility and opportunities for optimization
as different policies may be appropriate along different modes.

\section{Performance Metrics}
In this section, we identify certain fundamental metrics that determine the 
computational load and communication volume incurred by the HOOI procedure, 
under a given distribution scheme $(\pi_1, \pi_2, \ldots, \pi_N)$. 
The HOOI procedure consists of three components, TTM, SVD and factor matrix transfer, of which the first two involve computation
and the latter two involve communication.
We analyze the efficacy of a scheme along each mode $n$ separately. The cumulative performance across all modes
can be computed via suitable aggregation.

\subsection{Computational Load}
We generalize the notations $\calE^p$ and $\Rnp$ to multi-policy schemes in a natural manner.
Let $\calE_n^p$ denote the the set of elements owned by processor $p$ along mode $n$, i.e., $\calE_n^p = \{e : \pi_n(e) = p\}$.  
For a row-index $l\in [1, L_n]$, we say that $p$ {\em shares} $\slc_n^l$, 
if it owns at least one element from the slice with respect to the policy $\pi_n$, i.e., $\calE_n^p\cap \slc_n^l\neq \emptyset$. 
Let $\Rnp$ denote the number of slices shared by $p$ along mode $n$.

\paragraph{TTM Computation.}
The TTM component has to evaluate $|\calE|$ Kronecker products (one for each element). Thus, the TTM computation load is the same
for all distribution schemes. However, the policy $\pi_n$ may induce load imbalance by distributing the elements 
among the processors in a non-uniform manner. The TTM load imbalance along mode $n$ is captured by our first metric:
\begin{eqnarray*}
	\mbox{Metric 1:} && \Enmax = \max_p |\calE_n^p|.
\end{eqnarray*}
The optimal value the metric is the average $\avgE$, which can be achieved by uniformly distributing
the elements. 

\paragraph{SVD Computation.}
For any matrix-vector product associated with a Lanczos query, 
each processor $p$ incurs a computational load 
of $\Rnp\cdot \khat_n$, the size of its local copy. 
Let $Q_n$ be the number of queries raised by the Lanczos procedure.
Summed across all queries and all processors, the total oracle load is given by $Q_n\cdot \khat_n \cdot \sum_p \Rnp$.
Since $Q_n$ and $\khat_n$ are the same for all policies,
the oracle load along mode $n$ is captured by our second metric: 
\begin{eqnarray*}
	\mbox{Metric 2:} && \Rnsum = \sum_p \Rnp.
\end{eqnarray*}
Similarly, the load imbalance within the oracle computation along mode $n$ is captured by our third metric:
\begin{eqnarray*}
	\mbox{Metric 3:} && \Rnmax = \max_p~\Rnp.
\end{eqnarray*}
The above discussion has omitted other computations (such as internal to the Lanczos algorithm) that are common across all schemes.

The metric $\Rnsum$ is the aggregate number of times the slices are shared, across all processors.
We say that a slice is {\em good}, if it is shared by only one processor; otherwise, the slice is said be {\em bad}.
If a slice $S$ is good, then the corresponding row is non-empty only in the local copy of the processor sharing $S$,
whereas for a bad slice, the row becomes non-empty in multiple local copies.
Thus, bad slices lead to redundancy in the penultimate matrix and SVD computation, and result in higher load. 
An optimal policy is to assign each slice in its entirety to a single processor, thereby making all the slices
good. This would yield the optimal value of $\Rnsum = L_n$.
In addition, if the processors are assigned an equal number of slices, we get the optimal value of $\avgL$ on the metric $\Rnmax$.
For example, in Figure \ref{fig:framework}, each mode-$1$ slice is shared by two processors, leading to $\Rnsum = 6$, which 
results in a two-factor increase in the load compared to the setting where all slices are good.

\subsection{Communication Volume}
\paragraph{SVD Communication.}
For each query raised by the Lanczos procedure, the SVD component aggregates the local answers to get the global answers.
Consider the first product $\xout = \mnuf{Z}\cdot \xin$.
For each row-index $l\in [1, L_n]$, each processor sharing $\slc_n^l$, except the owner $\sigma_n(l)$,
sends one unit each of data (a real number) to the owner, where the data gets accumulated.
The number of units is the same as the number of processors sharing the slice minus one.
Summed across all row-indices, the communication volume per matrix-vector product is $\Rnsum - L_n$.
The product $\yin \cdot \mnuf{Z}$ can also be shown to incur the same volume.
Summed across all the $Q_n$ queries, the oracle communication volume along mode $n$
is $Q_n\cdot (\Rnsum - L_n)$. Since $Q_n$ and $L_n$ are constants across schemes, 
we can measure the oracle volume by $\Rnsum$, the same metric that determines the oracle load.
As before, we have omitted other communications that are common across the schemes.

\paragraph{Factor Matrix Transfer.}
Each row $\matF_n[l,:]$ of the factor matrix must be communicated to all the processors 
that would require the row for TTM computation in the next HOOI invocation.
In the case of uni-policy schemes, a processor requires the row, if it shares $\slc_n^l$. 
Excluding the owner, the number of processors is $\slc_n^l - 1$.
Each row consists of $K_n$ entries. Summed across all rows, the total communication volume is $K_n \cdot (\Rnsum - L_n)$ units. 
Thus, for uni-policy schemes, the factor matrix transfer volume is also determined by the parameter $\Rnsum$.

The case of multi-policy schemes is more intricate:
a processor requires the row $\matF_n[l,:]$, if it owns an element $e\in \slc_n^l$ with respect to 
any of the $(N-1)$ policies, excluding $\pi_n$.
(i.e., there exists an element $e\in \slc_n^l$ and a mode $j\neq n$ such that $\pi_j(e) = p$).
Hence, for multi-policy schemes, the factor matrix volume cannot be determined 
from our metrics. We shall measure the volume empirically.

The above metrics measure the efficacy of a scheme along a given mode $n$. The cumulative performance across all modes
can be computed via suitable aggregation, considering the mode-specific factors such as 
the number of queries $Q_n$ and the core length $K_n$.

\paragraph{Summary.}
We identified the following parameters influencing the HOOI execution time and the associated metrics:
TTM load balance ($\Enmax$), SVD computation load and communication volume ($\Rnsum$),
SVD load balance ($\Rnmax$). The factor matrix communication volume is also determined by $\Rnsum$ for uni-policy schemes.

\subsection{Computation vs Communication}
\label{sec:comp-comm}
It is useful to understand the breakup of the HOOI execution time in terms of computation and communication time.
Here, we present an intuitive comparison, taking as example $3$-D tensors and uniform core size of $K_1, K_2, K_3 = K$.
Along mode $n$, the TTM and SVD components involve $m\cdot K^2$ and 
$Q_n \cdot K^2 \cdot \Rnsum$ units of computation (FLOPs), respectively ($m$ is the number of non-zero elements).
In accordance with SLEPc \cite{slepc}, our implementation
of the Lanczos method involves $2\cdot K$ iterations, resulting in $Q_n = 4\cdot K$ queries. 
On the other hand, the SVD component and the factor matrix transfer components incur $Q_n \cdot (\Rnsum - L_n)$ 
and $K_n (\Rnsum - L_n)$ units of communication, respectively.
We can observe that the amount of computation is significantly larger than communication, especially
for distribution schemes with low redundancy ($\Rnsum$ being close to $L_n$).
The intuition is confirmed by our experimental evaluation, which shows that the computation time
is dominant even for the multi-policy schemes considered in the study.

\subsection{CP vs Tucker Decomposition}
It is of interest to compare the CP and the Tucker decompositions.
Both follow the ALS paradigm and the elements get distributed using a suitable scheme.  
The factor matrix transfer step is similar, but the other operations are significantly different.
As an illustration, consider a $3$-D tensor of size $L\times L \times L$ and core of size $K \times K \times K$.
The main computation in CP is the matricized tensor times Khatri-Rao product (MTTKRP):
for each element, the operation computes the Hadamard product of two $K$-length vectors ($O(K)$ FLOPs).
The corresponding operation in HOOI is the Kronecker product ($O(K^2)$ FLOPs).
In addition, HOOI computes the SVD of a large penultimate matrix of size $L\times K^2$.
As a result, computation time is the dominant factor in HOOI. In the case of CP, load balance and communication volume are important. 
In the case of Tucker, load balance ($\Enmax$ and $\Rnmax)$ and SVD redundancy ($\Rnsum$) are important,
and it is crucial to have low SVD redundancy, perhaps even at the cost of higher communication.
Hence, design considerations for distribution schemes for Tucker become different.
Schemes that work best for CP may not work as well for Tucker, and vice versa.

\section{Prior Distribution Schemes}
\label{sec:prior}
In this section, we discuss prior schemes proposed in the context of Tucker decomposition \cite{ucar-icpp},
as well the related CP decomposition \cite{dms-ipdps}. The schemes can be categorized in to three types.

\begin{figure}[t]
	\centering
	\includegraphics[width=2in]{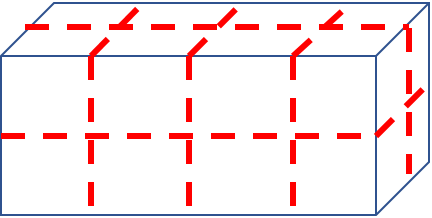}
	\caption{
	Example grid for $P = 16$. The grid is $4 \times 2 \times 2$.
	}
	\label{fig:grid}
\end{figure}

\paragraph{Coarse Grained Schemes.}
These are multi-policy schemes. Along each mode $n$, the policy $\pi_n$ is constructed by assigning each slice
in its entirety (all its elements) to a suitably chosen processor. 
All the slices are good and the metric $\Rnsum$ (capturing SVD load) attains the optimal value of $L_n$.
However, these schemes typically perform poorly on the metric $\Enmax$ (capturing TTM load balance),
since real-life tensors tend to have slices that are much larger than the average $\avgE$.
The imbalance can be somewhat mitigated via careful slice assignement strategies such as below \cite{dms-ipdps}:
arrange the mode-$n$ slices in a random order and allocate contiguous blocks of slices to the processors.
Other strategies similar in spirit have been proposed in prior work \cite{ucar-icpp, defacto, coarse2}.
We denote the above scheme as {\coarseg}.
Even with the above heuristics, coarse grained schemes tend to incur high TTM load imbalance.

\paragraph{Fine Grained Schemes.}
These are uni-policy schemes that address the TTM load imbalance by assigning individual elements, rather than entire slices.
Here, the key issue is to ensure that each slice is shared by few processors so that SVD redundancy is low.
Towards that goal, building on prior work \cite{hyper1}, 
Kaya and {\ucar} \cite{ucar-icpp} devised a fine grained scheme via 
reduction to hypergraph partitioning, a well-studied NP-hard problem.
The idea is to construct a hypergraph by taking the elements as vertices and the slices (along all modes) as hyperedges.
Then, we construct a (uni-policy) $\pi$ by finding a balanced min-cut partitioning.
The formulation models both the metrics $\Emax$ and $\Rnsum$.
Though the scheme achieves good performance on the HOOI execution time,
the time taken for hypergraph partitioning is significantly higher than the HOOI execution time (single invocation).
Consequently, the scheme is used offline.  We denote the schme as {\hyperg}.

\paragraph{Medium Grained Scheme.}
For CP decomposition, Smith and Karypis \cite{dms-ipdps} proposed a lightweight,
a medium-grained scheme that strikes a tradeoff between the above to schemes.
The idea is to factorize the number of processors $P$ in a suitable manner $P=q_1 \times q_2 \times \cdots \times q_N$
and overlay a processor grid of the above size over the tensor.
Then, each sub-tensor is assigned to a processor.
The indices along each mode are randomly permuted to offset any skew in element distribution within the input tensor.
The choice of the processor grid is crucial in determining the performance.
Along mode $n$, each slice can be shared by up to $P/q_n$ processors in the worst case
and so, $q_n$ is fixed in proportion to $L_n$.
We denote the scheme as {\mediumg}.
Figure \ref{fig:grid} provides an example grid.
%In practice, the scheme performs well in terms of TTM load balance and it is competitive with hypergraph based scheme 
%on the computational load and communication volume associated with the SVD component.

\paragraph{Row-Index Mapping.}
As in prior work \cite{dms-ipdps}, we fix the row-index mapping $\sigma_n$ as follows.
For each row $\matF_n[l,-]$, the owner is selected to be one among the processors sharing the slice $\slc_n^l$,
taking into account communication load balance arising in the SVD and the factor matrix transfer operations.

\section{Distribution Scheme {\Lite}}
Among the prior schemes, {\coarseg} is optimal on the metric $\Rnsum$, whereas {\mediumg} and {\hyperg} are superior on the metric $\Enmax$.
Uni-policy schmes (such as {\mediumg} and {\hyperg}) suffer from higher SVD redundancy,
since they try to construct a single policy that can perform well on all the modes simultaneously.
Multi-policy schemes can optimize the process better by constructing $N$ distribution policies, 
each customized for the computation along a single mode.
In this section, we present a lightweight, multi-policy scheme called {\Lite},
which is provably near-optimal on all the three metrics $\Enmax$, $\Rnsum$ and $\Rnmax$,
resulting in better computation time.
Though the scheme may incur higher communication volume,
it achieves better HOOI time, since computation time is the dominant factor.

\begin{figure}[t]
	\center
	\includegraphics[width=6in]{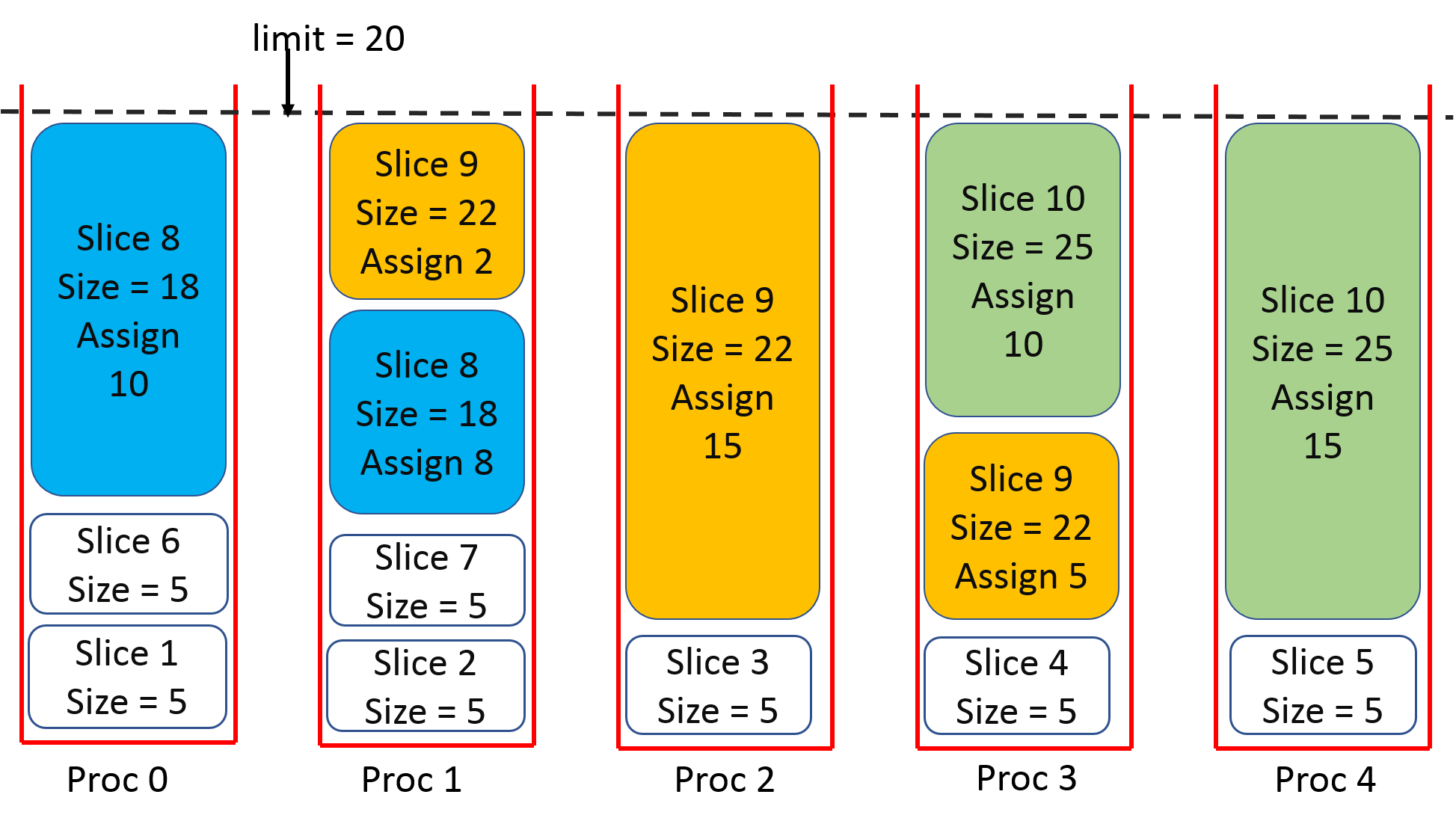}
	\caption{Illustration of {\Lite}.
		Here, $|\calE| = 100$ and $P = 5$. The limit is $\avgE = 20$. 
		We have ten slices with sizes in the sorted order as: 5, 5, 5, 5, 5, 5, 5, 18, 22 and 25. 
		Seven slices get processed in a round robin fashion in the first stage; if we assign the eighth slice to processor
		2, it would get $23$ elements, violating the limit. For the three slices processed in the second stage,
		the number of elements assigned to each processor is also shown.
	}
	\label{fig:bins}
\end{figure}

\subsection{{\Lite} Scheme}
The intuition behind {\Lite} is drawn from coarse grained schemes, which have optimal $\Rnsum$, but suffer 
from TTM load imbalance, because the elements may get distributed in a non-uniform manner.
We can attempt to address the issue by carefully assigning the slices so that the maximum number of elements
received by the processors (i.e., $\Enmax$) is minimized. The problem is the same as the 
classical makespan minimization on identical parallel machines \cite{vazirani}:
assume that the processors are machines and each slice $S$ is a task with execution time equal to $|S|$;
we wish to assign the tasks to the machines so that the makespan (overall completion time) is minimized.
The problem is NP-hard and heuristics with approximation guarantees are known.
For instance, a well-known heuristic is the best processor fit (BPF) procedure: scan the slices and assign each slice to the currently least loaded processor.
The above heuristic is guaranteed to output a solution within factor $2$ of the optimal solution.

There are two issues with the above approaches. The first is that the tensor may have very large slices,
in which case, even the optimal assignment of slices would incur high value of $\Enmax$ and TTM load imbalance.
Secondly, the processors may receive an uneven number of slices, leading to high value of $\Rnmax$ and SVD load imbalance.
In designing {\Lite}, we address the first issue by sharing the large slices among multiple processors,
and show that the second issue can be addressed by sorting the slices
in the increasing order of their sizes.

We next describe the {\Lite} distribution scheme along mode $n$.
Imagine that the processors are bins that need to be filled by the elements.
We wish to achieve the optimal value on the metric $\Enmax$, given by the average $\avgE$.
We consider the above value to be a hard limit on the number of elements that can be added to a bin.
The construction first sorts the mode-$n$ slices by their cardinalities and proceeds in two stages.

In the first stage, we consider the slices in the increasing order of cardinalities and assign them to the processors
in a round-robin fashion. We stop the process, if assigning a slice to the current bin would make it violate the limit.
At this point, we move to the second stage. 
The remaining slices are large in size and we fill the bins to their limit by 
sharing the large slices among multiple processors. To this effect, we scan the large slices and the bins concurrently.
If the whole of the current slice can be assigned to the current bin
without violating the limit, we do so and move to the next slice. Otherwise, we arbitrarily select elements 
from the current slice and add to the current bin till the limit is reached, and then move to the next bin.
Thus, the elements of each large slice get assigned to a contiguous set of processors.

In the above scheme, towards achieving fast tensor distribution time,
we sort the slices using the parallel sample-sort algorithm \cite{sample-sort}, a divide-and-conquer strategy similar to quicksort.
Given an array of keys, the idea is to use random sampling to derive a set of $P$ keys called splitters.
We then partition the array into $P$ buckets based on the splitters and let each processor sort a bucket independently.
Figure \ref{fig:bins} provides an illustration.
A pseudocode is given in Figure \ref{fig:lite};
it outputs the set of elements $\calE_n^p$ assigned to each processor $p$.

\begin{figure}[t]
\begin{center}
\begin{boxedminipage}{\hsize}
\begin{small}
\begin{tabbing}
xx\=xx\=xx\=xx\=xx\=xx\=xx\=\kill
$L \leftarrow L_n$\\
Sort $\slc_n^1, \slc_n^2, \ldots, \slc_n^L$ in increasing order of cardinality.\\
Let $S_1, S_2, \ldots, S_L$ be the slices in sorted order.\\
$\lmt \leftarrow \avgE$.\\
For all $l\in [1, L]$, $\calE_n^p \leftarrow \emptyset$.\\
\textbf{Stage 1:}\\
$p \leftarrow 0$\\
For $t = 1, 2, 3, \ldots$\\
\> If ($|\calE_n^p \cup S_t| > \lmt$) then GOTO Stage 2\\
\> else\\
\> \> Assign all elements of $S_t$ to $p$: $\calE_n^p \leftarrow \calE_n^p \cup S_t$.\\
\> \> $p \leftarrow (p+1) \bmod P$ 
\\
\textbf{Stage 2:}\\
$p \leftarrow 0$\\
while($p < P$)\\
\> $g \leftarrow \lmt - |\calE_n^p|$. \quad\quad // gap with respect to $\lmt$\\
\> If($|S_t| \leq g$) then\\
\> \> Assign all elements of $S_t$ to $p$: $\calE_n^p \leftarrow \calE_n^p \cup S_t$.\\
\> \> $t \leftarrow t+1$ \quad\quad\quad // Move to next slice\\
\> else\\
\> \> $X\leftarrow$ Select any $g$ elements from $S_t$.\\
\> \> Assign selected elements to $p$: $\calE_n^p \leftarrow \calE_n^p \cup X$\\
\> \> Remove selected elements from $S_t$: $S_t \leftarrow S_t\setminus X$\\
\> \> $p \leftarrow p+1$ \quad\quad // Move to next processor
\end{tabbing}
\end{small}
\end{boxedminipage}
\end{center}
\caption{$\Lite$: Distribution along mode $n$}
\label{fig:lite}
\end{figure}

\subsection{Performance Guarantee and Discussion}
\begin{theorem} 
	\label{thm:main}
	For the scheme {\Lite}, along any mode $n$,
	\begin{enumerate}
		\item $\Enmax \leq \avgE$.
		\item $\Rnsum \leq L_n + P$.
		\item $\Rnmax \leq \avgL + 2$.
	\end{enumerate}
\end{theorem}

A proof sketch is provided in Section \ref{sec:proof}.
At a high level, the first metric is explicitly ensured by setting the hard limit.
Regarding the other two metrics, all the slices processed in the first stage are good
and the round-robin process implies that every bin receives the same number of slices.
We shall argue that at most $P$ slices can remain in the second stage
and that each processor can share at most two of them.

The theorem shows that the scheme is optimal on the metric $\Emax$ and achieves perfect TTM load balance.
Recall that the optimal values for the metrics $\Rnsum$ and $\Rnmax$ are $L_n$ and $\avgL$, respectively.
On the above two metrics, {\Lite} is away from optimality only by additive factors of $P$ and $2$, respectively.
Within the SVD component, the communication volume per matrix-vector product is given by $\Rnsum - L_n$
and so, the scheme incurs only $P$ units of communication per matrix-vector product.
Thus, the scheme is near-optimal on the computational load, load balanace and communication volume associated with the SVD component.
We note that the scheme may incur overall higher communication volume, due to higher factor matrix data transfer.
However, as shown in our experimental study,
{\Lite} outperforms the prior schemes on the overall HOOI execution time,
since the computation time is the dominant factor.

We next briefly compare the memory requirements of {\Lite} with that of prior schemes.
Being a multi-policy scheme, {\Lite} needs to store $N$ copies of the input tensor, one along each mode.
However, due to low SVD redundancy, the scheme requires lesser space for storing the penultimate matrices.
Consequently, as shown by our experimental evaluation, 
{\Lite} performs better or comparable to the prior schemes in terms of the memory requirements.

A recent work on the CP decomposition \cite{smith-ipdps17} tries to handle the large slices
by uniformly distributing the elements of slices larger than a heuristically determined threshold.
Our algorithm {\Lite} solves the isuse by using a principled approach yielding near-optimal bounds.

\subsection{Proof of Theorem \ref{thm:main}}
\label{sec:proof}
Part (1) is readily true, since the scheme explicitly ensures that the number of elements assigned to any processor is at most $\avgE$. 
We next prove part (2) of the theorem. Let $\that$ denote the iteration in which the procedure made the exit to the second stage.
Let $\calS_1$ and $\calS_2$ denote the slices processed in the first and the second stages, respectively;
namely, $\calS_1 = \{S_1, S_2,\ldots, S_{\that -1}\}$ and $\calS_2 = \{S_{\that}, S_{\that+1}, \ldots, S_L\}$, where $L = L_n$.

By the construction of the second stage, each slice $S\in \calS_2$ is assigned to a set of contiguous processors.
We call the first among these processors as the {\em head} of $S$ and the others are said to form the {\em tail} of $S$.
As an illustration, in Figure \ref{fig:bins}, for slice 9, processor 1 is the head and the processors 2 and 3 form the tail.
Observe that any processor can participate in the tail of at most one slice.

Let $\share(\calS_1)$ denote the aggregate number of times the slices from $\calS_1$ are shared; define $\calS_2$ similarly.
The slices in $\calS_1$ are all good and so, $\share(\calS_1) = |\calS_1|$.
The quantity $\share(\calS_2)$ is same as the number of times the processors act as the heads plus the number of times
they participate in tails. The first quantity is $|\calS_2|$, since every slice has a single head. 
The second quantity is at most $P$, since as we observed earlier, every processor participates in the tail of at most one slice.
Therefore, $\share(\calS_2) \leq |\calS_2| + P$. Put together, we get that $\Rnsum$ is at most $L_n + P$,
proving part (2) of the theorem. Moreover, since every slice is shared by at least one processor, the above result 
also implies that there can be at most $P$ bad slices. 

%We derive part (4) of the theorem as a corollary of part (2).
%Let $\calG$ and $\calB$ denote the set of good and bad slices.
%Every slice in $\calG$ is shared exactly once and each slice in $\calB$ is shared at least twice.
%The quantity $\Rnsum$ is the number of times the slices are shared
%and hence, $\Rnsum \geq |\calG| + 2|\calB|$.  The total number of slices is $L_n = |\calG| + |\calB|$.
%Thus, $\Rnsum \geq L_n + |\calB|$. Therefore, part (2) implies that $|\calB| \leq P$.

We next prove part (3) of the theorem by showing that for any processor $p$, $\Rnp \leq \avgL + 2$.
The first stage assigns the slices in a round-robin fashion and so, the number of slices from $\calS_1$ assigned to $p$ is at most $\lceil |\calS_1|/P \rceil\leq \avgL$.
Regarding slices from $\calS_2$, an important issue is that while the slices in $\calS_1$ are good, those in $\calS_2$ can be potentially good or bad.
We say that a slice $S$ is {\em ugly}, if $S\in \calS_2$ and it is good.
The ugly slices pose a difficulty: it is hypothetically possible that a large number of ugly slices get assigned to the processor $p$ in the second stage,
leading to a high value of $\Rnp$ and load imbalance in the oracle computation. 
We eliminate the possibility by proving that ugly slices do not exist.

Towards that goal, we  first argue that the first stage 
follows the best processor fit strategy: namely, in any iteration $t$, the slice $S_t$ gets assigned to the processor having the least number of elements. 
Below, we formalize and the prove the claim.  For $t \geq 1$, let $p_t$ denote the processor that receives the slice $S_t$ in iteration $t$, i.e., $p_t = (t\mod P)$.
For an iteration $t$ and processor $p$, let $h_t(p)$ denote the number of elements assigned to $p$ till the beginning of iteration $t$ 
(not including the assignment made during the iteration $t$); thus $h_1(p) = 0$, for all $p$. 

\begin{lemma}
\label{lem:one}
For any iteration $t\geq 1$, we have that $h_t(p_t) \leq h_t(p)$, for all $p$. 
\end{lemma}
\proof
We prove the lemma by establishing a stronger statement that the number of elements assigned to the processors
are in ascending order in a cyclic manner starting with $p_t$, and the difference between the largest and the smallest assignments 
does not exceed $|S_t|$.  The proof goes via induction and the strengthening helps in the induction step.
For any processor $p$, we write `$p\oplus 1$' and `$p\ominus 1$' to mean `$(p+1)\mod P$' and `$(p-1)\mod P$', respectively.

{\it Claim:}
For any iteration $t\geq 1$: (a) for any $p\neq p_t$, $h_t(p) \geq h_t(p\ominus 1)$; (b) $h_t(p_t\ominus 1) - h_t(p_t) \leq |S_t|$.

The lemma follows from part (a) of the claim.  We prove the claim by induction. 
The claim is trivially true for the base case $t=1$, since all the processors are empty to start with.
Assume that the claim is true for $t$ and we prove it for $t+1$.
By the induction hypothesis, at the beginning of iteration $t$, $p_t$ has the least and $(p_t\ominus 1)$ has the largest number of elements,
and the difference is at most $|S_t|$. During the iteration $t$, $S_t$ gets assigned to $p_t$.
So, $p_t$ becomes the processor with the largest number of elements and $p\oplus 1$ becomes the processor with the least number of elements.
Barring $p_t$, the ordering among the other processors remains the same. Thus, for any $p\neq (p\oplus 1)$, $h_{t+1}(p) \geq h_{t+1}(p\ominus 1)$.
Since $p_{t+1} = p\oplus 1$, we have proved part (a).
Regarding part (b), $p_t$ has lesser number of elements than $p\oplus 1$ in the beginning of iteration $t$
and receives the slice $S_t$. Hence, $h_{t+1}(p_t) - h_{t+1}(p\oplus 1) \leq |S_t| \leq |S_{t+1}|$;
rephrased, $h_{t+1}(p_{t+1} \ominus 1) - h_{t+1}(p_{t+1}) \leq |S_{t+1}|$. Part (b) is proved.
\qed

For a processor $p$, let $\hhat(p)$ denote the number of elements assigned to $p$ at the end
of the first stage and let $\ghat(p) = \avgE - \hhat(p)$ denote the gap to the limit. 
We next use Lemma \ref{lem:one} to argue that any slice $S\in \calS_2$ is too big to fit the gap of any processor.

\begin{lemma}
\label{lem:two}
For any $S\in \calS_2$ and processor $p$, $|S| > \ghat(p)$.
\end{lemma}
\proof
Let $\phat = \that \mod P$. The exit to the second stage implies that $|S_{\that}| + \hhat(\phat) > \avgE$, or equivalently $|S_{\that}| > \ghat(\phat)$.
Since the slices are sorted, $S_{\that}$ has the least cardinality among the slices in $\calS_2$.
On the other hand, Lemma \ref{lem:one} implies that $\phat$ has the least number of elements at the beginning of iteration $\that$,
or equivalently, $\phat$ has the largest gap. The lemma is proved.
\qed

The above lemma shows that second stage cannot assign any slice from $\calS_2$ in its entirety to a single processor;
namely, ugly slices do not exist. Thus, all the slices in $\calS_2$ are shared by at least two processors
and hence, any processor $p$ can act as the head of at most slice from $\calS_2$.
We observed earlier that $p$ can participate in the tail of at most one slice from $\calS_2$.
Therefore, $p$ can share at most two slices from $\calS_2$.
Since $p$ shares at most $\avgL$ from $\calS_1$, we get that $\Rnp \leq \avgL + 2$. 
Part (3) of the theorem is proved.

\section{Experimental Evaluation}
In this section, we evaluate the performance of {\Lite} and the prior schemes {\coarseg}, {\mediumg} and {\hyperg},
(the best known scheme under each category) on HOOI execution time, distribution time, memory usage and related statistics.

\begin{figure}[t]
	\centering
\begin{tabular}{l|llllll}
	Tensor & $L_1$ & $L_2$ & $L_3$ & $L_4$ & nnz & Sparsity \\
\hline
{\tt delicious} &  532K & 17.2M & 2.4M  & 1.4K & 140M & $4.2\times 10^{-15}$\\
{\tt enron}     & 6K   & 5K    & 244K  & 1K   & 54M & $5.4\times 10^{-9}$\\
{\tt flickr}    &  319K & 28M   & 1.6M  & 731  & 112M & $1.0 \times 10^{-14}$\\  
{\tt nell1}     & 2.9M & 2.1M  & 25.4M & -    & 143M   & $9.1\times 10^{-13}$\\                                                             
{\tt nell2}     & 12K  & 9K    & 28K   & -    & 77M  & $2.4\times 10^{-5}$\\
\hline
{\tt amazon}    & 4.8M & 1.7M  & 1.8 M & - & 1.7B & $1.1\times 10^{-10}$\\
{\tt patents}   & 46 & 239 K & 239 & - & 3.5B & $1.3\times 10^{-3}$\\
{\tt reddit}    & 8.2M & 176K & 8.1M & - & 4.6B & $3.9\times 10^{-10}$
\end{tabular}
\caption{Tensor datasets}
\label{fig:datasets}
\end{figure}

\subsection{Experimental Setup}
\paragraph{System.}
The experiments were conducted on a cluster of Power-8 nodes (20 cores, $256$ GB, $4$ GHz)
connected via InfiniBand in a fat-tree topology. 
We launch 16 MPI ranks per node, each mapped to a core. 
We use $2$ to $32$ nodes, leading to $32$ to $512$ MPI ranks.

{\it Tensor Datasets: }
The dataset consists of eight real-world tensors drawn from the FROSTT repository \cite{frostt}
that represent data from NLP datasets, social bookmarking and rating services. 
Of the eight tensors, five are medium-sized with at least $50$ million elements
and the other three are big tensors with more than billion elements.
For each tensor, Figure \ref{fig:datasets} shows the length along each mode, the number of non-zero elements (nnz)
and the sparsity (ratio of the number of non-zero elements to the total size of the tensor).
While the first three are $4$-dimensional, the others are $3$-dimensional.

{\it Implementation: }
Our implementation is based on MPI. We use the iterative Lanczos bidiagonalization method \cite{slepc} for SVD operation.
In accordance with SLEPc \cite{slepc}, we set the number of Lanczos iterations to be $2K$, where $K$ is the number
of singular vectors requested.  
We use ATLAS 3.10.1 for dense linear algebra.
The compiler used is gcc 4.8.5. 

For the {\hyperg} scheme, we obtained the hypergraph partitioning using the parallel Zoltan library \cite{zoltan}, used in prior work \cite{dms-ipdps} as well.
We could not obtain the partitioning for the three big tensors using the library.
So, we consider the scheme only on the medium-sized tensors.

%We evaluated two hypergraph partioners used in prior work \cite{ucar-icpp,dms-ipdps}: PATOH \cite{patoh}, a sequential library and Zoltan \cite{zoltan}, a parallel library.
%In the experiments below, we use the partitions produced by Zoltan, 
%since the library is faster in terms of distribution time
%and for our dataset, it resulted in HOOI execution times that is better or close to those of PATOH.
%We could not obtain the partitioning for the three big tensors using the library.
%So, we consider the scheme only on the medium-sized tensors.

{\it Core Size:}
The HOOI time is dependent on the size of the core tensor.
As in prior work \cite{ucar-icpp,kolda-icdm,baskaran}, we use a uniform core length of $K_n = K$ for all modes $n$
and set $K=10$ in all the experiments, except one, where we study the effect of increasing the core size.

\begin{figure}
\centering
\begin{tabular}{c}
\begin{tabular}{cc}
\includegraphics[width=2.2in]{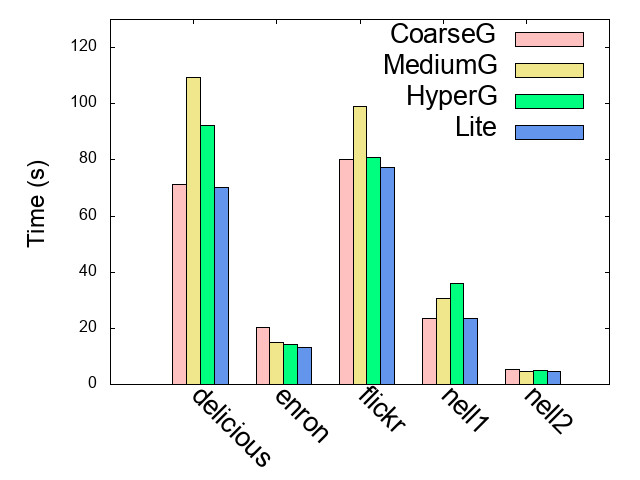}
&
\includegraphics[width=2.2in]{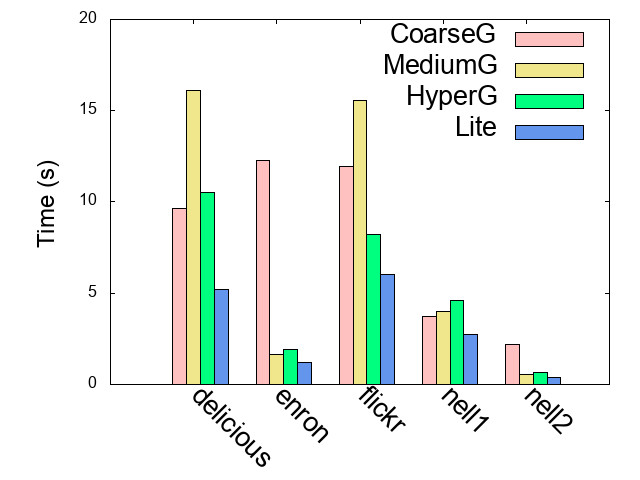}
\\
(a) $K = 10$, ranks = $32$ 
&
(b) $K = 10$, ranks = $512$ 
\end{tabular}
\\
\includegraphics[width=2.2in]{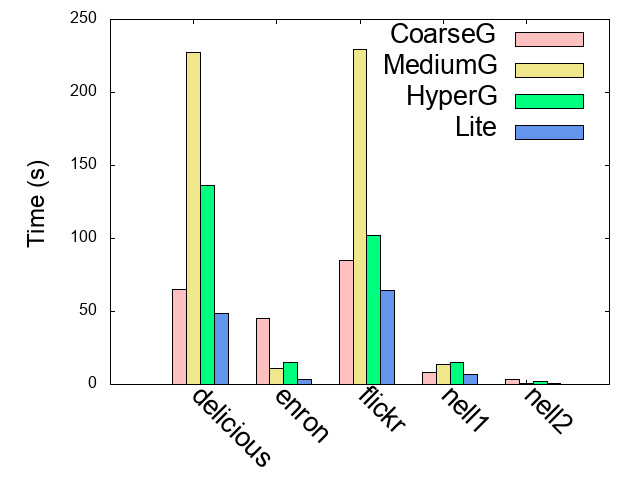}
\\
(c) $K = 20$, ranks = $512$ 
\end{tabular}
\caption{
HOOI execution time comparison on the medium-sized tensors.
}
\label{fig:basic}
\end{figure}

\begin{figure}
\centering
\begin{tabular}{c}
\begin{tabular}{cc}
\includegraphics[width=2.2in]{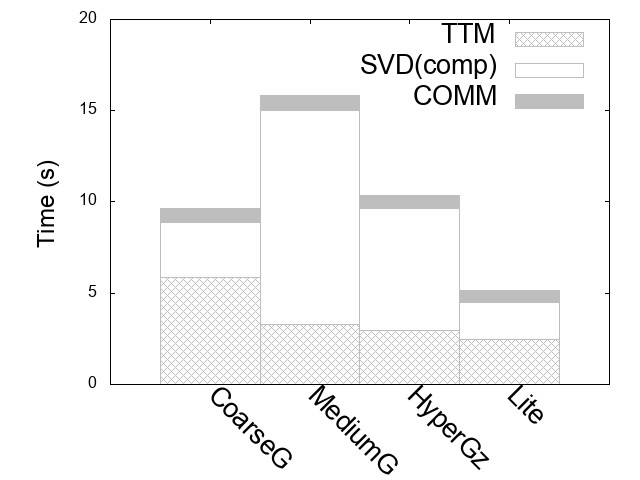}
&
\includegraphics[width=2.2in]{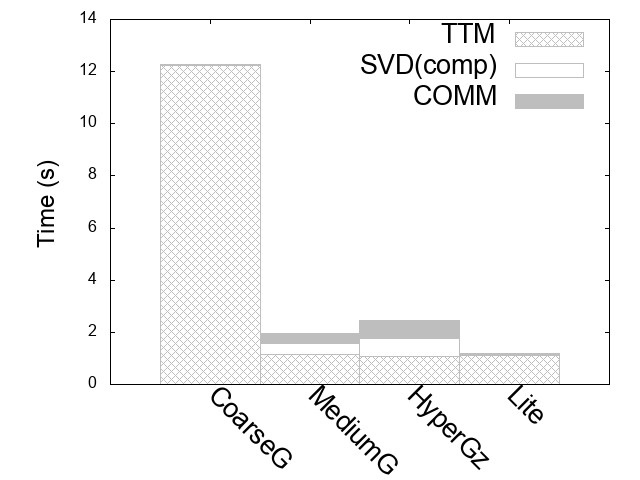}
\\
(a) {\tt delicious} & (b) {\tt enron}
\end{tabular}
\\
\includegraphics[width=2.2in]{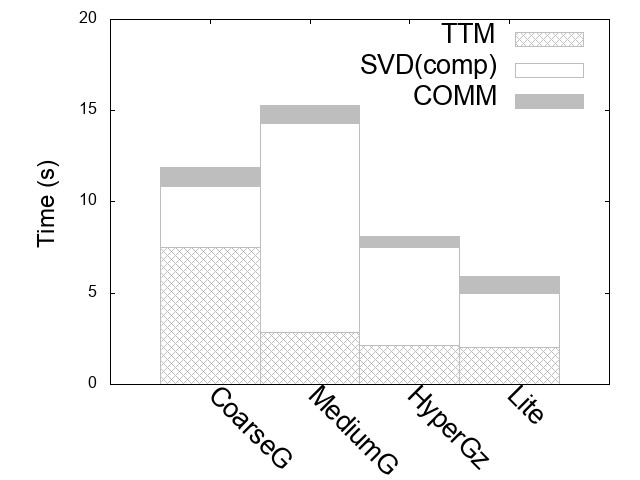}
\\
(c) {\tt flickr}
\end{tabular}
\caption{HOOI execution time breakup for $K=10$, number of ranks $512$}
\label{fig:breakup}
\end{figure}

\subsection{HOOI Execution Time}
We first compare the different schemes on the HOOI execution time (single invocation) on the medium-size tensors;
the big tensors are considered separately later in the section.
We consider three different configurations. Setting $K=10$, the first two configurations consider the smallest ($32$) and the largest ($512$) number of ranks in our setup.
The third configuration studies the effect of increasing the core size, and sets $K=20$ and number ranks as $512$. 

\paragraph{HOOI Execution Time.}
The execution times are shown in Figure \ref{fig:basic}.
Among the prior schemes, the scheme offering the least execution time varies across the test cases
and overall {\hyperg} has better performance.
We can see that {\Lite} offers the best performance on all the datasets and configurations.
It outperforms {\coarseg}, {\mediumg} and {\hyperg} by factors upto $12$x, $4.5$x and $4.1$x, respectively. 
Compared to the best prior scheme in each test case, the performance improves by a factor of up to $3$x,
with the performance gain increasing with increase in number of ranks and core size.

Towards understanding the above phenomenon, 
we analyze the HOOI components and the underlying metrics.  
For this purpose, we use the second configuration ($K=10$ and ranks$=512$)
and the first three tensors as illustrative example.

\paragraph{Time Breakup.}
Figure \ref{fig:breakup} provides the breakup of HOOI execution time
in terms of TTM and SVD computation time, and the total communication time (SVD plus the factor matrix transfer).
We can see that the computation time dominates the overall execution time.
While {\coarseg} is better on SVD, {\mediumg} and {\hyperg} are better on TTM computation.
{\Lite} performs well on both the components.

\begin{figure}
\centering
\begin{tabular}{c}
\begin{tabular}{cc}
\includegraphics[width=2.2in]{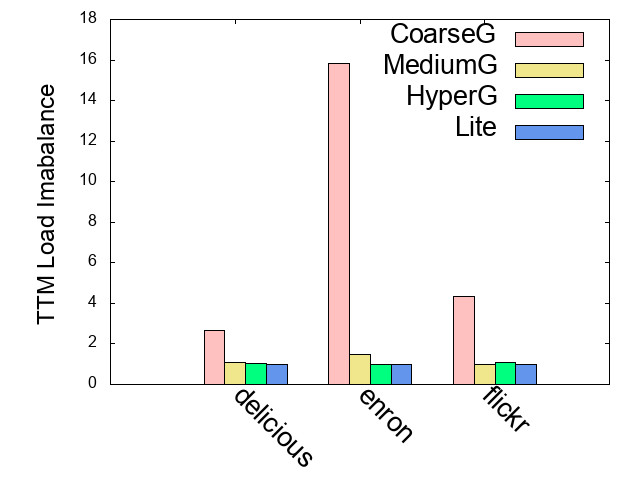}
&
\includegraphics[width=2.2in]{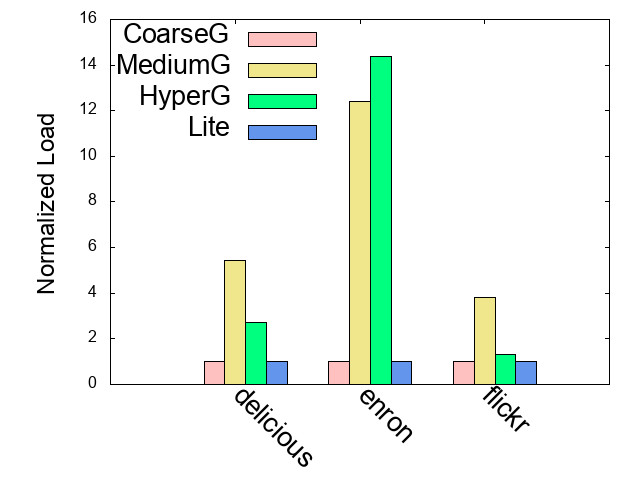}
\\
(a) TTM Load Imbalance & (b) SVD Computational Load 
\end{tabular}
\\
\includegraphics[width=2.2in]{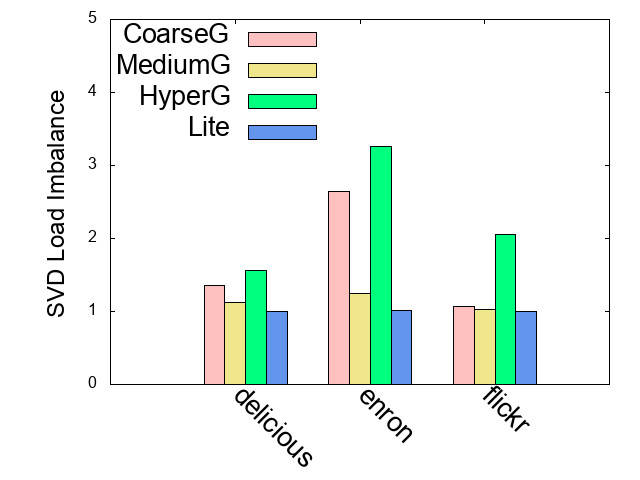}
\\
(c) SVD Load Imbalance
\end{tabular}
\caption{Analysis of computation time parameters at $K=10$ and ranks = $512$}
\label{fig:params}
\end{figure}

\begin{figure}
\centering
\begin{tabular}{c}
\begin{tabular}{cc}
\includegraphics[width=2.2in]{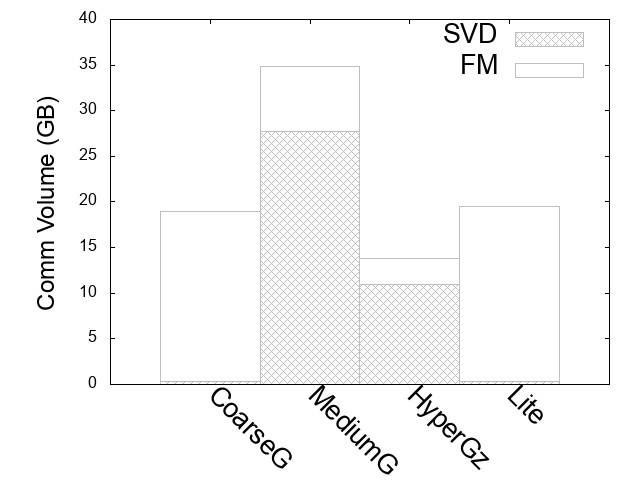}
&
\includegraphics[width=2.2in]{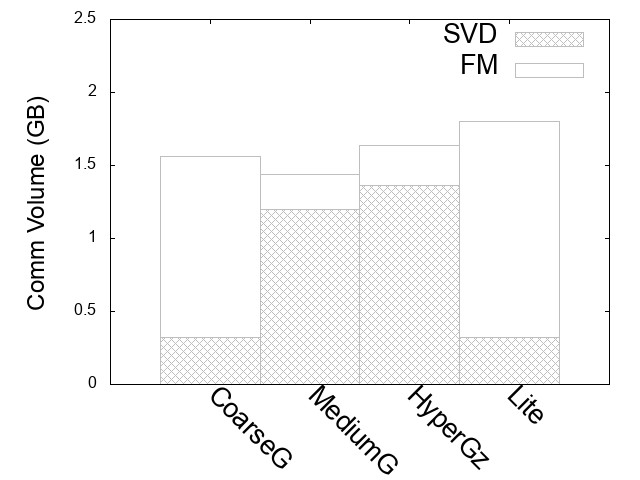}
\\
(a) {\tt delicious} & (b) {\tt enron}
\end{tabular}
\\
\includegraphics[width=2.2in]{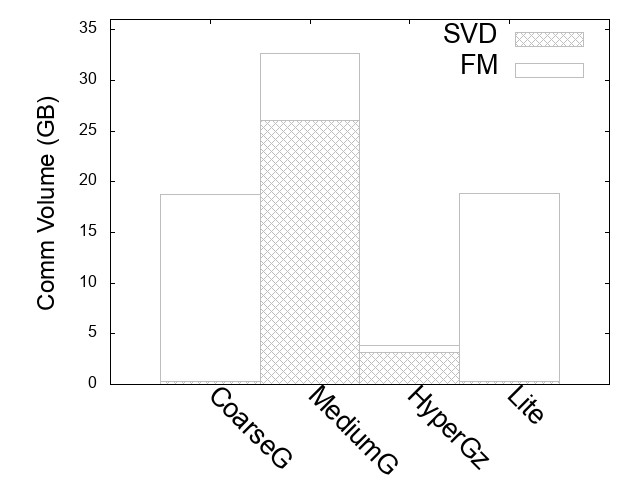}
\\
(c) {\tt flickr}
\end{tabular}
\caption{Communication volume for $K=10$, number of ranks $512$}
\label{fig:comm}
\end{figure}

\paragraph{Computation  Metrics.}
The above behavior can be understood by analyzing the underlying metrics, presented in Figure \ref{fig:params}.
The computation time is determined by the TTM load balance, SVD computational load and load imbalance.
We measure the computational load (FLOPs) by taking the aggregate along all the modes.
The load balance is given by the ratio of maximum to the average across the processors, with the optimal value being one.

From Figure \ref{fig:params} (a), we can see that {\mediumg}, {\hyperg} and {\Lite} achieve near-perefct TTM load balance. 
The {\coarseg} scheme performs poorly, because it assigns entire slices to the processors
and as a result, the processors receiving large slices induce load imbalance.
For instance, {\tt Enron} has $54$M elements yielding an average of $105$K elements per processor at $512$ ranks,
but the tensor has slices of size $5$M elements. 
The above example shows that the severe load imbalance cannot be mitigated
even by careful slice assignment mechanisms such as the best processor fit.

The optimal SVD load is attained when each slice is owned by a single processor. 
We measure the redundancy in the SVD computation by normalizing the load with respect to the optimal value; the normalized load in shown in Figure \ref{fig:params} (b).
Being uni-policy schemes, {\mediumg} and {\hyperg} have to contend with computations across multiple modes simultaneously, leading to high redundancy.
Recall that under {\mediumg}, each mode-$n$ slice can be shared by up to $P/q_n$ processors in the worst case.
Though not reaching the worst case bound, we can see that the redundancy is high under {\mediumg}, resulting in higher 
HOOI execution time.
In contrast, $\coarseg$ achieves the optimal redundancy of one unit, since all the slices are good under the scheme.
Being near-optimal on the metric $\Rnsum$, $\Lite$ attains redundancy close to one.

Regarding SVD load balance, we can see from Figure \ref{fig:params} (c) that 
{\Lite} performs well, since it is guaranteed to be near-optimal on the metric $\Rnmax$.
The {\mediumg} scheme also performs well.

\begin{figure}
\centering
\begin{small}
\begin{tabular}{c|ccccc|c}
Time(s)   & {\coarseg} 	& {\mediumg}  	& {\Lite} 	\\ \hline
amazon    & 89.0   	& 13.1    	& 8.6 		\\
patents   & 96.4   	& 15.5    	& 14.2  		\\
reddit    & 232.1   	& 23.6    	& 21.6 
\end{tabular}
\end{small}
\caption{HOOI execution time on the big tensors}
\label{fig:big_tensors}
\end{figure}

\paragraph{Communication Volume.}
We observed that the computation time dominates the HOOI time.
Here, we analyze the small communication time by considering the communication volume.
Figure \ref{fig:comm} shows the breakup in terms of the SVD and the factor matrix transfer (FM) components.
The SVD component involves communication during oracle query answering and other communication 
that are common across the schemes. The oracle communication volume along mode $n$ is given by $\Rnsum - L_n$.
The metric $\Rnsum$ is $L_n$ for the {\coarseg} and close to $L_n$ for {\Lite}.
Hence, the two schemes incur little SVD communication, but being multi-policy schemes, they have higher FM volume. 
In contrast, {\mediumg} and {\hyperg} incur lesser FM volume, but higher SVD volume.
The {\hyperg} scheme achieves good tradeoff and performs the best on the overall communication volume.
Nevertheless, {\Lite} outperforms {\hyperg} on HOOI execution time, since the computation time is the dominant factor.

\paragraph{Big Tensors.}
We next consider the three big tensors at $512$ ranks and $K=10$.
The smallest of these tensors has $1.7$ billion elements and the aggregate cardinalities of 
the hyperedges is three times bigger. Given the large size of the resulting hypergraph,
we could not obtain the hypergraph partitioning on these tensors using the two libraries.
The execution times for the other three schemes are shown in Figure \ref{fig:big_tensors}.
The number of non-zero elements in these tensors is much larger compared to their dimension lengths
and as a result, the TTM computation time dominates.
The {\coarseg} scheme performs poorly due to TTM load imbalance, since these tensors have very large slices.
In contrast, by the virtue of good TTM load balance, the other two schemes perform well.
We can see that {\Lite} achieves the best execution time on all the three tensors and
outperforms {\mediumg} by a factor of up to $1.5$x.

\begin{figure}
\centering
\begin{tabular}{c}
\begin{tabular}{c|cccc}
Speedup		&{\coarseg}	&{\mediumg}	&{\hyperg}	&{\Lite}\\ \hline
delicious	&7.4		&6.8		&8.8		&13.4 \\
enron		&1.7		&9.0		&7.4		&11.1 \\
flickr		&6.7		&6.4		&9.8		&12.9 \\
nell1		&6.4		&7.6		&7.9		&8.6 \\
nell2		&2.4		&8.4		&7.5		&12.2 \\
amazon		&1.8		&11.0		&x		&13.5 \\
patents		&2.7		&14.5		&x		&15.5 \\
reddit		&1.8		&14.2		&x		&14.6
\end{tabular}
\\
\\
(a) Speedup from $32$ to $512$ ranks
\\
\\
\includegraphics[width=3in]{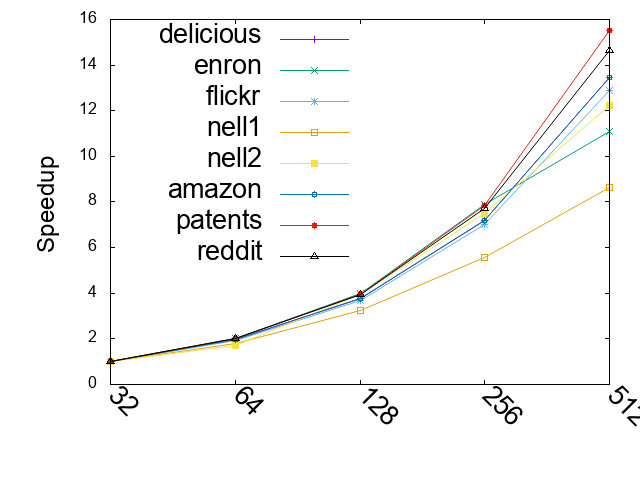}
\\
(b) {\Lite} Strong scaling
\end{tabular}
\caption{Scaling study}
\label{fig:scaling}
\end{figure}

\subsection{Scaling, Distribution Time and Memory}
\paragraph{Strong scaling.}
We studied the scaling behavior of the HOOI procedure under the different schemes by varying 
the number of ranks from $32$ to $512$.
The speedup results are reported in Figure \ref{fig:scaling}.
As the number of ranks increases, the average TTM load decreases,
but the sizes of the large slices remain the same. 
As a result,  $\coarseg$ suffers from severe load imbalance and scales poorly. 
The other schmes scale comparatively better. 
The figure also includes the scaling for all ranks from $32$ to $512$ ranks under the {\Lite} scheme.
We see that the scheme exhibits the best scaling behavior:
as against an ideal value of $16$, {\Lite} achieves speedup in the range $8.6 - 15.5$x, which translates to a scaling efficiency of $55 - 97$\%.

\begin{figure}
\centering
\begin{tabular}{c|cccc|c}
	Time (s)		&{\coarseg}	&{\mediumg}	&{\hyperg}	&Lite & HOOI\\ \hline
delicious & 6.8   	& 9.3    	& 345   	& 3.9  		& 5.2  \\
enron     & 0.1   	& 0.08    	& 125    	& 0.1  		& 1.1  \\
flickr    & 10.9   	& 14.0    	& 203   	& 5.5 		& 6.0  \\
nell1     & 10.5   	& 13.9    	& 356   	& 6.2 		& 2.7  \\
nell2     & 0.07   	& 0.05    	& 91    	& 0.07  	& 0.3  \\
amazon    & 2.9   	& 5.5    	& x      	& 2.5 		& 8.7  \\
patents   & 3.2   	& 0.9    	& x      	& 2.0  		& 14.2 \\
reddit    & 7.8   	& 11.6    	& x      	& 5.7  		& 21.6
\end{tabular}
\caption{Distriution time}
\label{fig:distr}
\end{figure}

\paragraph{Distribution Time.} 
We next evaluate the schemes on the time taken for distributing the input tensor under the configuration of $K=10$ and $512$ ranks; see Figure \ref{fig:distr}.
We implemented the three lightweight schemes in parallel as part of the HOOI procedure and 
obtained the {\hyperg} partitioning by executing the Zoltan library in parallel in an offline fashion.
For the {\Lite} scheme, the distribution time refers to the time spent in executing a parallel implementation of the procedure given in Figure \ref{fig:lite}.
The HOOI execution time the under {\Lite} scheme is also included for the sake of comparison.
We can see that the distribution times of the three lightweight schemes are lesser or comparable to the HOOI execution time,
whereas {\hyperg} takes significantly higher time.

\begin{figure}
\center
\begin{tabular}{c}
\begin{tabular}{c|cccc}
	Memory (MB)   	& {\coarseg} 	& {\mediumg}  	& {\hyperg} 	& {\Lite}\\ \hline
delicious	&383	&1748	&881	&385\\
enron		&17	&53	&61	&17\\
flickr		&533	&1813	&625	&533\\
nell1		&112	&197	&151	&113\\
nell2		&12	&6	&19	&12\\
amazon		&348	&371	&x	&350\\
patents		&445	&158	&x	&447\\
reddit		&814	&543	&x	&812\\
\end{tabular}
\\
(a) Total memory 
\\
\begin{tabular}{cc}
\includegraphics[width=3in]{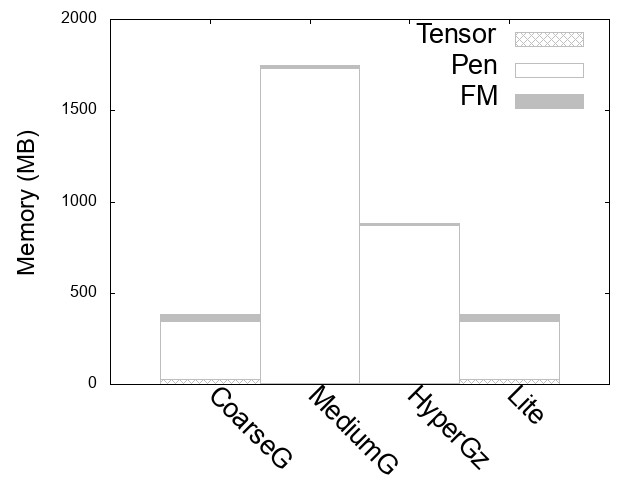} &
\includegraphics[width=3in]{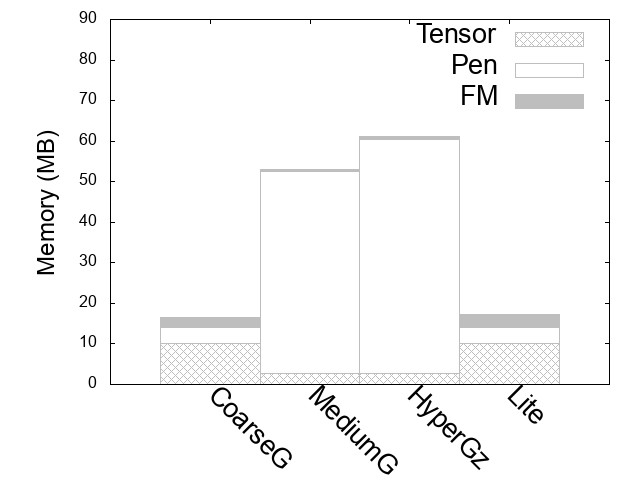}
\\
(b) delicious & (c) enron
\end{tabular}
\\
\includegraphics[width=3in]{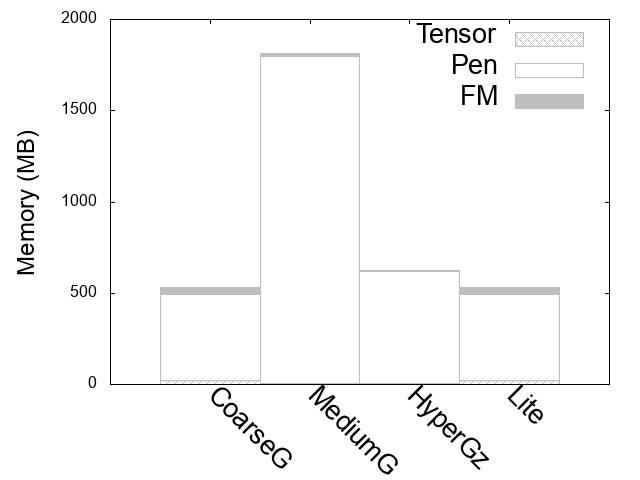}
\\
(d) flickr 
\end{tabular}
\caption{Memory Usage - Avergae memory (MB) per rank}
\label{fig:mem}
\end{figure}

\paragraph{Memory Usage.}
Th procedure needs to store the input tensor, the penultimate matrices and the factor matrices.
We next evalute the above memory requirement under the different schemes.
The results are shown in Figure \ref{fig:mem}. 
For the first three tensors, the figure also includes the breakup in terms of the three components.
Being multi-policy schemes, {\coarseg} and {\Lite} store $N$ copies of the input tensor,
and they do not actively minimize the factor matrix storage.
However, due to low redundancy, they take lesser amount of space to store the penultimate matrices.
In contrast, {\mediumg} and {\hyperg} store only a single copy of the tensor and are better on factor matrix storage,
but due to higher redudancy, they take more space for storing the penultimate matrices.
The size of the penultimate matrices get larger with increase in the number of dimension, whereas
the space needed for storing the tensor increaes with the increase in density of the input tensor.
We see that {\Lite} and {\coarseg} require nearly the same amount of memory.
They outperform {\mediumg} and {\hyperg} on the first three tensors, which are four dimensional.
The other tensors are three dimensioanl, with the three large tensors being relatively denser.
As a result, {\mediumg} performs better on these tensors.
Overall, we can see that {\Lite} is better or comparable to {\mediumg} and {\hyperg} on the overall memory requirement.

\section{Conclusions and Future Work}
In this paper, we proposed an improved lightweight distribution scheme for the Tucker decomposition of sparse tensors.
The scheme is provably near-optimal on certain fundamental metrics that determine the HOOI execution time.
Our experimental evaluation demonstrates that the scheme performs well in terms of distribution time and outperforms prior schemes 
on the HOOI execution time by a factor of upto $3$x.  
We identify two avenues for future work related to shared memory systems.
While the new scheme is near-optimal on the TTM and the SVD components, it does not explicitly optimize 
the factor matrix communication volume. 
Since the above communication does not arise in shared memory systems, 
the new scheme may provide optimal strategies for partitioning work among the threads. 
Conversely, recent work on shared memory systems \cite{europar} has shown that
the TTM computational load can be reduced using the compressed sparse fiber representation.
The strategy may be useful in optimizing the TTM computations local to the processors.

\paragraph{Acknowledgments.}
We thank the anonymous referees of earlier versions of the paper for useful comments and suggestions. 
They helped in improving the paper significantly.

\bibliographystyle{plain}
\bibliography{main}

\appendix
\section{Reformulation}
Here, we describe how to compute the rows of the penultimate matrix.
Let $u_1, u_2, \ldots, u_r$ be a sequence vectors of length $s_1, s_2, \ldots, s_r$, respectively.
The Kronecker (or outer) product of the sequence is an $r$-dimensional tensor
of size $s_1\times s_2\times \ldots \times s_r$, wherein the element with coordinate
$(c_1, c_2, \ldots, c_r)$ takes the value $\Pi_{j=1}^r u_j[c_j]$.
We represent the tensor as a vector of length $s_1\cdot s_2\cdot \ldots \cdot s_r$ 
by arranging the elements in a lexicographic order.
Namely, the above value is placed in position $\sum_j c_j \left(\Pi_{i < j} \ell_i \right)$.

For an element $e$ with coordinate $(l_1, l_2, \ldots, l_n\ldots, l_N)$ and value $\val(e)$,
let $\kron_n(e)$ denote the vector yielded by the Kronecker product of the
following rows (vectors) of the factor matrices:
\[
\matF_n[l_1, :], \ldots, \matF_{n-1}[l_{n-1}, :], \matF_{n+1}[l_{n+1}, :], \ldots, \matF_N[l_N, :].
\]
Define $\contr_n(e)=\val(e)\cdot \kron_n(e)$.
For any $l\in [1, L_n]$, the row $\mnuf{Z}[l,:]$ is given by the summation shown in (\ref{eqn:reform}).

\end{document}